\documentclass[pre,preprint,showpacs,eqsecnum]{revtex4}
\usepackage{bm}
\usepackage{graphicx}
\usepackage{amsmath,amssymb}

\begin{document}
\title{Limiting distributions of continuous-time random walks
with superheavy-tailed waiting times}
\author{S.~I.~Denisov$^{1}$}
\email{denisov@sumdu.edu.ua}
\author{Yu.~S.~Bystrik$^{1}$}
\author{H.~Kantz$^{2}$}
\affiliation{$^{1}$Sumy State University, Rimsky-Korsakov Street 2,
UA-40007 Sumy, Ukraine\\
$^{2}$Max-Planck-Institut f\"{u}r Physik komplexer Systeme,
N\"{o}thnitzer Stra{\ss}e 38, D-01187 Dresden, Germany}


\begin{abstract}
We study the long-time behavior of the scaled walker (particle)
position associated with decoupled continuous-time random walk which
is characterized by superheavy-tailed distribution of waiting times
and asymmetric heavy-tailed distribution of jump lengths. Both the
scaling function and the corresponding limiting probability density
are determined for all admissible values of tail indexes describing
the jump distribution. To analytically investigate the limiting
density function, we derive a number of different representations of
this function and, by this way, establish its main properties. We
also develop an efficient numerical method for computing the limiting
probability density and compare our analytical and numerical results.
\end{abstract}

\maketitle

\section{INTRODUCTION}
\label{Intr}

The continuous-time random walks (CTRWs), i.e., cumulative jump
processes which are characterized by a joint probability density of
waiting time and jump length, play a significant role in many areas
of science. The reason is that a large variety of physical,
biological and other systems are often modeled by two random
variables that can be interpreted as the waiting time between
successive transitions (jumps) of the system into new states and the
transition measure (jump length). For example, the CTRW model and its
modifications can be used to describe anomalous diffusion and
transport in disordered media \cite{SM, MK, AH, KRS}, human mobility
\cite{BHG, SKWB}, financial \cite{SGM, MMW, Scal, MNPW} and seismic
\cite{HS, PAG} data.

According to the theory of CTRWs \cite{MW} (see also Refs.~\cite{MK,
AH}), the probability density $P(x,t)$ of the particle position
$X(t)$ depends only on the joint probability density of waiting time
and jump length. Unfortunately, even in the simplest (decoupled) case
when the joint density is a product of waiting-time density $p(\tau)$
and jump-length density $w(\xi)$, the representation of $P(x,t)$ in
terms of special functions is known in a few cases \cite{KZ, Bar,
GPSS}. In contrast, the long-time behavior of $P(x,t)$ that
determines the diffusion and transport properties of walking
particles is studied analytically in much more detail \cite{Tun, SKW,
WWH, Kot, MS}. Specifically, the asymptotic behavior of the scaled
particle position $Y(t) = a(t)X(t)$ is investigated in
Ref.~\cite{Kot}. In this work, the scaling function $a(t)$ and the
distribution function of $Y(t)$ at $t \to \infty$ are obtained for
all waiting-time and jump densities characterized by finite second
moments or heavy tails.

A special case of CTRWs with \textit{superheavy-tailed} distributions
of waiting time was first considered in \cite{HW}. Because all
fractional moments of these distributions are infinite, they can be
used to model extremely anomalous behavior of  systems and processes
as, e.g., iterated maps \cite{DrKl}, ultraslow kinetics \cite{CKS},
superslow diffusion \cite{DK1}, and Langevin dynamics \cite{DKH,
DK2}. For the CTRWs characterized by \textit{arbitrary}
superheavy-tailed distributions of waiting time, the scaling function
$a(t)$ and the corresponding limiting probability density
$\mathcal{P}(y)$ of $Y(\infty)$ have already been found for the jump
densities having finite second moments \cite{DK3} and symmetric heavy
tails \cite{DYBKL}. In the present work, we report a comprehensive
theoretical and numerical studies of the long-time behavior of the
reference CTRWs in the general case of \textit{asymmetric} jump
densities characterized by heavy tails.

The paper is organized as follows. In Sec.~\ref{Mod}, we describe the
CTRW model, formulate the problem of the limiting probability density
and list some previously obtained results. In Sec.~\ref{Lim}, we find
the scaling function $a(t)$ and the limiting probability density
$\mathcal{P}(y)$ in terms of the inverse Fourier transform for all
possible cases. A number of alternative representations of the
limiting density and its main properties are obtained in
Sec.~\ref{rep}. Here, we derive $\mathcal{P}(y)$ in terms of (i) the
inverse Mellin transform, (ii) the Laplace transform, (iii) the Fox
$H$ function, and (iiii) the series expansion. Using the series and
Laplace representations of the limiting probability density, in
Sec.~\ref{beh} we determine its short- and long-distance behavior. In
Sec.~\ref{Num}, we develop a method for the numerical evaluation of
$\mathcal{P}(y)$ and compare the analytical and numerical results.
Our findings are summarized in Sec.~\ref{Concl}. Finally, a short
derivation of the fractional equation for $\mathcal{P}(y)$ is
presented in the Appendix.

\section{MODEL AND BACKGROUND}
\label{Mod}

One of the main statistical characteristics of a CTRW is the
probability density $P(x,t)$ of the particle position $X(t)$. This
quantity is completely determined by the joint probability
 density $\Psi(\xi,\tau)$ of waiting times $\tau_{n}$ ($\tau_{n} \geq
0$), i.e., times between successive jumps, and jump lengths $\xi_{n}$
($-\infty<\xi_{n}< \infty$). The random variables in the sets
$\{\tau_ {n}\}$ and $\{\xi_{n}\}$ are assumed to be independent and
identically distributed with the probability densities $p(\tau)$ and
$w(\xi)$, respectively. In the case of decoupled CTRWs, when the sets
$\{\tau_{n}\}$ and $\{\xi_{n}\}$ are independent of each other,
$\Psi(\xi,\tau) = w(\xi)p(\tau)$ and the probability density $P(x,t)$
in Fourier-Laplace space is given by the Montroll-Weiss equation
\cite{MW}
\begin{equation}
    P_{ks} = \frac{1-p_{s}}{s(1-p_{s}w_{k})}.
\label{M-Weq}
\end{equation}
Here, $w_{k} = \mathcal{F} \{w(x)\} = \int_{-\infty}^{\infty} dx
e^{ikx} w(x)$ with $-\infty <k< \infty$ is the Fourier transform of
$w(\xi)$, $p_{s} = \mathcal{L} \{p(t)\} = \int_{0}^{\infty} dt
e^{-st} p(t)$ with $\mathrm{Re} s>0$ is the Laplace transform of $p(
\tau)$, and $P_{ks} = \mathcal{F} \{ \mathcal{L} \{P(x,t)\}\}$.

Representing Eq.~(\ref{M-Weq}) in the form
\begin{equation}
    P_{ks} = \frac{1-p_{s}}{s} + \frac{(1-p_{s})p_{s}w_{k}}
    {s(1-p_{s}w_{k})}
\label{M-Weq2}
\end{equation}
and taking the inverse Fourier transform [defined as $\mathcal{F}^
{-1} \{f_{k}\} = f(x) = (2\pi)^{-1} \int_{- \infty}^{ \infty} dk
e^{-ikx} f_{k}$] of Eq.~(\ref{M-Weq2}), one gets
\begin{equation}
    P_{s}(x) = \frac{1-p_{s}}{s}\, \delta(x) +
    \frac{(1-p_{s})p_{s}}{s}\, \mathcal{F}^{-1}
    \bigg\{\frac{w_{k}}{1-p_{s}w_{k}} \bigg\}
\label{P(x)s}
\end{equation}
with $\delta(x)$ being the Dirac $\delta$ function. Then, applying
the inverse Laplace transform [defined as $\mathcal{L} ^{-1}
\{g_{s}\} = g(t) = (2\pi i)^{-1} \int_{c- i\infty}^{c+ i\infty} ds
e^{st} g_{s}$, $c$ is a real number that exceeds the real parts of
all singularities of $g_{s}$] to Eq.~(\ref{P(x)s}), for the
probability density of the particle position we obtain
\begin{equation}
    P(x,t) = V(t) \delta(x) + \mathcal{L}^{-1}
    \bigg\{\! \frac{(1-p_{s})
    p_{s}}{s}\mathcal{F}^{-1} \bigg\{\frac{w_{k}}{1-p_{s}
    w_{k}}\bigg\}\! \bigg\},
\label{P(x,t)}
\end{equation}
where
\begin{equation}
    V(t)= \mathcal{L}^{-1} \bigg\{ \frac{1-p_{s}}{s} \bigg\}
    = \int_{t}^{\infty} d\tau p(\tau)
\label{defV}
\end{equation}
is the survival probability, i.e., the probability that a walking
particle remains at the initial state $X(0)=0$ up to time $t$.
According to the definition (\ref{defV}), this probability satisfies
the conditions $V(t) \to 0$ as $t \to \infty$ and $V(t) \to 1$ as $t
\to 0$.

Since there are no boundary conditions keeping a walking particle
inside a finite region, the condition $P(x,t) \to 0$ as $t \to
\infty$ must hold for all $x$. The fact that $P(x,t)$ vanishes in the
long-time limit suggests to introduce the scaled particle position
$Y(t) = a(t) X(t)$, where $a(t)$ is a positive scaling function, and
study, instead of $P(x,t)$, the asymptotic behavior of the
probability density $\mathcal{P}(y,t) = P(y/a(t),t) /a(t)$ of $Y(t)$.
It is clear that if $a(t)$ at $t \to \infty$ tends to zero fast
enough, then the limiting probability density
\begin{equation}
    \mathcal{P}(y) = \lim_{t \to \infty} \frac{1}{a(t)}
    \,P \! \left( \frac{y}{a(t)}, t \right)
\label{limP}
\end{equation}
reduces to the degenerate density $\mathcal{P}(y) = \delta(y)$. In
contrast, if $a(t)$ at $t \to \infty$ grows or tends to zero slowly
enough, then $\mathcal{P}(y,t)$ vanishes in the long-time limit.
Finally, for a certain asymptotic behavior of $a(t)$ the limiting
probability density becomes nonvanishing and nondegenerate. The last
property is of particular interest because in this case the pair
$\mathcal{P} (y)$ and $a(t)$ determines the asymptotic behavior of
the probability density of the particle position, $P(x,t) \sim a(t)
\mathcal{P} (a(t)x)$ as $t \to \infty$. To avoid any
misunderstanding, we note that this statement is correct only if
$\mathcal{P}(y) \neq 0$; in those regions of $y$ where
$\mathcal{P}(y) = 0$ (for details, see below) the use of the limiting
probability density for determining the long-time behavior of
$P(x,t)$ becomes impractical.

The problem of finding the pairs $\mathcal{P}(y)$ and $a(t)$ has been
solved for all typical distributions of waiting times and jump
lengths characterized by both finite second moments and heavy tails
\cite{Kot}. Recently, we have partially solved this problem for a new
class of CTRWs with superheavy-tailed distributions of waiting times
\cite{DK3, DYBKL}. These distributions are characterized by the
following asymptotic behavior of the waiting-time density:
\begin{equation}
    p(\tau) \sim \frac{h(\tau)}{\tau} \quad (\tau \to \infty),
\label{p as}
\end{equation}
where $h(\tau)$ is a slowly varying function defined by the condition
$h(\mu \tau) \sim h(\tau)$ ($\tau \to \infty$) holding for all
$\mu>0$. Since $p(\tau)$ is normalized, $\int_{0} ^{\infty}d\tau
p(\tau) =1$, the function $h(\tau)$ must decrease in such a way that
$h(\tau) = o(1/\ln \tau)$ as $\tau \to \infty$. The main feature of
these densities is that their fractional moments $\int_{0}^{\infty}
d\tau \tau^{c} p(\tau)$ are infinite for all $c>0$. It has been shown
\cite{DK2} that if $p(\tau)$ is superheavy-tailed and $w(\xi)$ has a
finite second moment $l_{2} = \int_{-\infty} ^{\infty} d\xi \xi^{2}
w(\xi)$, then
\begin{equation}
    \mathcal{P}(y) = \frac{2-\delta_{l_{1},0}}{2}\, e^{-|y|}
    H(l_{1}y)
\label{P_finite}
\end{equation}
and
\begin{equation}
    a(t) \sim \left\{\!\! \begin{array}{cl}
    \sqrt{2V(t)/l_{2}},
    & l_{1} = 0
    \\ [6pt]
    V(t)/|l_{1}|,
    & l_{1} \neq 0
    \end{array}
    \right.
\label{a_finite}
\end{equation}
as $t \to \infty$. Here, $\delta_{a,b}$ and $H(x)$ are the Kronecker
$\delta$ ($\delta_{ a,b} =1$ if $a=b$ and $\delta_{a,b} =0$ if $a
\neq b$) and the Heaviside unit function [$H(x) =1$ if $x\geq 0$ and
$H(x) =0$ if $x<0$], respectively, and $l_{1} = \int_{-\infty}
^{\infty} d\xi \xi w(\xi)$ is the first moment of $w(\xi)$. We note
that if $l_{1} \neq 0$, then the limiting probability density is
one-sided: $\mathcal{P} (y)=0$ on that semi-axis of $y$ where
$l_{1}y<0$.

If the jump density is symmetric, $w(-\xi) = w(\xi)$, and has heavy
tails, then
\begin{equation}
    w(\xi) \sim \frac{u}{|\xi|^{1+ \alpha} } \quad
    (|\xi| \to \infty),
\label{w_sym}
\end{equation}
where $u>0$ and $\alpha \in(0,2]$ is the tail index. According to
\cite{DYBKL}, in this case the limiting probability density can be
represented in the forms
\begin{eqnarray}
    \mathcal{P}(y) \!\!&=&\!\! \frac{1}{\alpha}H_{2,3}^{2,1}
    \left[ |y| \Big |
        \begin{array}{lcl}
        (1-1/\alpha,1/\alpha), (1/2,1/2)\\
        (0,1),(1-1/\alpha,1/\alpha),(1/2,1/2)
        \end{array}
    \right]
    \nonumber\\ [4pt]
    \!\!&=&\!\!  \frac{1}{\pi} \int_{0}^{\infty}dx
    e^{-|y|x}\frac{\sin (\pi \alpha/2) x^{\alpha}}
    {1 + 2\cos (\pi \alpha/2) x^{\alpha} + x^{2\alpha}}
    \nonumber\\ [4pt]
    \!\!&=&\!\!\frac{1}{\pi} \int_{0}^{\infty}dx
    \frac{\cos (yx)}{1 + x^{\alpha}}
\label{P_heavy}
\end{eqnarray}
($H_{2,3}^{2,1}[\cdot]$ is a particular case of the Fox function) and
the corresponding scaling function is given by
\begin{equation}
    a(t) \sim \left\{\!\! \begin{array}{cl}
    \left( \frac{\Gamma(1 + \alpha)
    \sin (\pi \alpha/2)}{\pi u} V(t)
    \right)^{1/\alpha},
    & 0<\alpha<2
    \\ [8pt]
    \sqrt{\frac{2V(t)}{u\ln[1/V(t)]}},
    & \alpha=2
    \end{array}
    \right.
\label{a_heavy}
\end{equation}
($t \to \infty$) with $\Gamma(1 + \alpha)$ being the $\Gamma$
function. Interestingly, Eq.~(\ref{P_heavy}) at $\alpha = 2$ reduces
to Eq.~(\ref{P_finite}) with $l_{1} =0$. But since in this case
$l_{2}= \infty$, the scaling function in Eq.~(\ref{a_heavy}) differs
from that given in Eq.~(\ref{a_finite}). It should also be stressed
that if $p(\tau)$ is superheavy-tailed, then the survival probability
$V(t)$ is a slowly varying function \cite{DK2}. Therefore, in
accordance  with Eqs.~(\ref{a_finite}) and (\ref{a_heavy}), the
long-time evolution of $P(x,t)$ occurs very slowly.

In this paper, we will study analytically the long-time solutions of
the CTRWs characterized by both superheavy-tailed distributions of
waiting time, whose asymptotic behavior is described by Eqs.~(\ref{p
as}), and heavy-tailed distributions of jump length. The last
distributions are assumed to be asymmetric and can have one or two
heavy tails. Since the limiting probability density $\mathcal{P}(y)$
under certain conditions is determined by the heaviest tail (see
below), we consider, without loss of generality, the jump densities
with two heavy tails, whose asymptotic behavior is given by
\begin{equation}
    w(\xi) \sim \frac{u_{\pm}}{|\xi|^{1+ \alpha_{\pm}} }
    \quad (\xi \to \pm\infty)
\label{w_asymm}
\end{equation}
($u_{\pm}>0$, $\alpha_{\pm} \in (0,2]$). In contrast to
Ref.~\cite{DYBKL}, now we are concerned with the effects arising from
the asymmetry of $w(\xi)$. In addition, we are going to develop a
numerical method for the simulation of these CTRWs and apply it to
verify the theoretical predictions.

\section{LIMITING PROBABILITY DENSITIES AND
CORRESPONDING SCALING FUNCTIONS} \label{Lim}

\subsection{Inverse Fourier transform representation of
$\bm{\mathcal{P} (y)}$} \label{3a}

Under certain conditions (see below), the long-time behavior of the
probability density $P(x,t)$ is determined by the small-$s$ behavior
of the Laplace transform $P_{s}(x)$. In turn, according to
Eq.~(\ref{P(x)s}), the last behavior is governed by the  small-$s$
behavior of $p_{s}$.  Taking into account that
\begin{eqnarray}
    1-p_{s} \!\!&=&\!\!  \int_{0}^{\infty}dt (1-e^{-st})p(t)
    \nonumber\\ [4pt]
    \!\!&=&\!\! \int_{0}^{\infty}dq e^{-q}V(q/s)
\label{relat}
\end{eqnarray}
and using the fact that the survival probability $V(t)$ is a slowly
varying function, we obtain
\begin{equation}
    1 - p_{s} \sim V(1/s)
\label{as ps}
\end{equation}
as $s \to 0$, and Eq.~(\ref{P(x)s}) in this limit yields
\begin{eqnarray}
    P_{s}(x) \!\!&\sim&\!\! \frac{V(1/s)}{s}\, \delta(x) +
    \frac{V(1/s)[1-V(1/s)]}{2\pi s}
    \nonumber\\ [4pt]
    &&\!\! \times \int_{-\infty}^{\infty}dk
    \frac{e^{-ikx}}{V(1/s) + 1 -w_{k}}.
\label{P(x)s2}
\end{eqnarray}

Applying to $P_{s}(x)$ the Tauberian theorem \cite{Fel} [it states
that if the function $v(t)$ is ultimately monotonic and $v_{s} \sim
s^{-\gamma} L(1/s)$ ($0<\gamma< \infty$) as $s \to 0$, then $v(t)
\sim t^{\gamma-1} L(t)/ \Gamma( \gamma)$ as $t \to \infty$, where
$L(t)$ is a slowly varying function at infinity], from
Eq.~(\ref{P(x)s2}) one gets
\begin{eqnarray}
    P(x,t) \!\!&\sim&\!\! V(t)\delta(x) +
    \frac{V(t)[1-V(t)]}{2\pi}
    \nonumber\\ [4pt]
    &&\!\! \times \int_{-\infty}^{\infty}dk
    \frac{e^{-ikx}}{V(t) + 1 -w_{k}}
\label{P(x,t)2}
\end{eqnarray}
($t \to \infty$). With this result, we can represent the limiting
probability density (\ref{limP}) as the inverse Fourier transform
\begin{equation}
    \mathcal{P}(y) = \frac{1}{2\pi}\int_{-\infty}^{\infty}
    d\kappa \frac{e^{-i\kappa y}}{1 + \Phi(\kappa)},
\label{limP2}
\end{equation}
where
\begin{equation}
    \Phi(\kappa) = \lim_{t \to \infty}
    \frac{1 -w_{\kappa a(t)}}{V(t)}.
\label{defPhi}
\end{equation}
We remark that, since $\int_{-\infty}^{+\infty} dy e^{-i \kappa y} =
2\pi \delta(\kappa)$ and $\Phi(0)=0$, this probability density is
properly normalized: $\int_{-\infty}^{+\infty} dy \mathcal{P}(y) =1$.
It should be noted also that $\mathcal{P}(y)$ satisfies a simple
space-fractional equation (see the Appendix).

Below, using Eqs.~(\ref{limP2}) and (\ref{defPhi}), we determine the
limiting probability density and corresponding scaling function in
all possible situations.

\subsection{Jump densities with $\bm{l_{1} \neq 0}$}
\label{3b}

If the first moment $l_{1}$ of the jump density exists and is
non-zero, then directly from the relation $1 - w_{k} =
\int_{-\infty}^{\infty} (1 - e^{ikx})w(x)$ one finds $1 - w_{k} \sim
- il_{1}k$, and thus Eq.~(\ref{defPhi}) reduces to
\begin{equation}
    \Phi(\kappa) = -i \kappa\, \mathrm{sgn}(l_{1})
    \lim_{t \to \infty} \frac{|l_{1}| a(t)}{V(t)}
\label{Phi2}
\end{equation}
[$\mathrm{sgn}(x) = \pm 1$ if $x\gtrless 0$]. Choosing the asymptotic
behavior of the scaling function in the form
\begin{equation}
    a(t) \sim \frac{V(t)}{|l_{1}|}
\label{as_a2}
\end{equation}
($t \to \infty$), Eq.~(\ref{Phi2}) yields $\Phi(\kappa) = -i \kappa\,
\mathrm{sgn}(l_{1})$. Then, using this result, from Eq.~(\ref{limP2})
we obtain the one-sided exponential density
\begin{equation}
    \mathcal{P}(y) =  e^{-|y|} H(l_{1}y).
\label{limP3}
\end{equation}

This limiting probability density describes all CTRWs characterized
by superheavy-tailed distributions of waiting time and jump
distributions having non-zero first moments. It should be emphasized
that a class of these jump distributions contains both the
distributions with finite second moments, see Eq.~(\ref{P_finite}),
and the heavy-tailed distributions with $\alpha_{\pm} \in (1,2]$.

\subsection{Jump densities with  $\bm{\alpha \in (1,2)}$ and
$\bm{l_{1} =0}$} \label{3c}

If the first moment of $w(\xi)$ exists and equals zero, then, to find
the the asymptotic behavior of $1-w_{k}$ as $k \to 0$, it is
reasonable to use the following exact formula:
\begin{eqnarray}
    1-w_{k} \!\!&=&\!\! \frac{1}{|k|}
    \int_{0}^{\infty} dx (1-\cos x)
    w^{+}\! \bigg( \frac{x}{|k|} \bigg)
    \nonumber\\ [4pt]
    &&\!\! + \frac{i}{k}
    \int_{0}^{\infty} dx (x-\sin x)
    w^{-}\! \bigg( \frac{x}{|k|} \bigg),
\label{w_k1}
\end{eqnarray}
where
\begin{equation}
    w^{\pm}(\xi) = w(\xi) \pm w(-\xi).
\label{w_pm}
\end{equation}
Let us introduce the notation
\begin{equation}
    \alpha = \min\{\alpha_{+}, \alpha_{-}\}
\label{min}
\end{equation}
and consider the case with $\alpha \in (1,2)$. Then, using the
standard integrals \cite{PBM}
\begin{equation}
    \int_{0}^{\infty} dx \frac{1-\cos x}{x^{1+\nu}} =
    \frac{\pi}{2\Gamma(1+\nu) \sin(\pi \nu/2)}
\label{int1}
\end{equation}
($0<\nu<2$) and
\begin{equation}
    \int_{0}^{\infty} dx \frac{x-\sin x}{x^{1+\nu}} =
    -\frac{\pi}{2\Gamma(1+\nu) \cos(\pi \nu/2)}
\label{int2}
\end{equation}
($1<\nu<2$), it can be shown from Eq.~(\ref{w_k1}) that
\begin{equation}
    1-w_{k} \sim  q|k|^{\alpha} - i \mathrm{sgn}(k)
    r|k|^{\alpha}
\label{as_w1}
\end{equation}
($k \to 0$), where
\begin{equation}
    q = \frac{\pi}{2\Gamma(1+\alpha) \sin(\pi \alpha/2)}
    (u_{+}\delta_{\alpha \alpha_{+}} + u_{-}\delta_{\alpha
    \alpha_{-}})
\label{q}
\end{equation}
and
\begin{equation}
    r = \frac{\pi}{2\Gamma(1+\alpha) \cos(\pi \alpha/2)}
    (u_{+}\delta_{\alpha \alpha_{+}} - u_{-}\delta_{\alpha
    \alpha_{-}}).
\label{r}
\end{equation}
It is worth to emphasize that Eqs.~(\ref{as_w1})--(\ref{r}) hold for
all admissible values of the largest tail index $\alpha_{\mathrm
{max}} = \max\{ \alpha_{+}, \alpha_{-} \}$, i.e., for $\alpha_{
\mathrm{max}} \in [\alpha,2]$.

Assuming that the long-time behavior of the scaling function is given
by
\begin{equation}
    a(t) \sim \left( \frac{V(t)}{\sqrt{q^{2} + r^{2}}}
    \right)^{1/\alpha}
\label{as_a3}
\end{equation}
($t \to \infty$), from Eqs.~(\ref{defPhi}) and (\ref{as_w1}) one
obtains
\begin{equation}
    \Phi(\kappa) = (\cos \varphi - i\, \mathrm{sgn}(\kappa)
    \sin \varphi )|\kappa|^{\alpha},
\label{Phi3}
\end{equation}
and the limiting probability density (\ref{limP2}) in this case can
be represented as follows:
\begin{equation}
    \mathcal{P}(y)\! = \frac{1}{\pi} \int_{0}^{\infty}\!\!
    d\kappa \frac{(1 + \cos \varphi \kappa^{\alpha})
    \cos(y\kappa) + \sin \varphi \kappa^{\alpha}
    \sin(y\kappa)}{1 + 2\cos\varphi \kappa^{\alpha}
    + \kappa^{2\alpha}}.
\label{limP4}
\end{equation}
Here,
\begin{equation}
    \cos \varphi = \frac{q}{\sqrt{q^{2} + r^{2}}},
    \quad \sin \varphi = \frac{r}{\sqrt{q^{2} + r^{2}}}
\label{def_phi}
\end{equation}
and
\begin{eqnarray}
    \sqrt{q^{2} + r^{2}} \!\!&=&\!\! \frac{\pi}
    {2\Gamma(1+\alpha) \sin(\pi \alpha/2)
    |\cos(\pi \alpha/2)|}
    \nonumber\\ [4pt]
    &&\!\! \times\, (2\cos(\pi \alpha)u_{+}u_{-}
    \delta_{\alpha \alpha_{+}} \delta_{\alpha
    \alpha_{-}}
    \nonumber\\ [4pt]
    &&\!\! +\, u_{+}^{2} \delta_{\alpha \alpha_{+}}\! +
    u_{-}^{2} \delta_{\alpha \alpha_{-}})^{1/2}.
\label{relat1}
\end{eqnarray}

The limiting probability density (\ref{limP4}) and the corresponding
scaling function (\ref{as_a3}) describe both symmetric and asymmetric
CTRWs. In particular, if $\alpha_{+} = \alpha_{-} = \alpha \in (1,2)$
and $u_{+} = u_{-} = u$, then $\varphi =0$,
\begin{equation}
    \sqrt{q^{2} + r^{2}} = \frac{\pi u}
    {\Gamma(1+\alpha) \sin(\pi \alpha/2)},
\label{relat2}
\end{equation}
and Eqs.~(\ref{limP4}) and (\ref{as_a3}) are reduced to
Eqs.~(\ref{P_heavy}) and (\ref{a_heavy}) describing the symmetric
CTRWs.

\subsection{Jump densities with $\bm{\alpha \in (0,1)}$}
\label{3d}

Since at $\alpha \in (0,1)$ the first moment of the probability
density $w(\xi)$ does not exist, in this case it is convenient to
use, instead of Eq.~(\ref{w_k1}), the following formula:
\begin{eqnarray}
    1-w_{k} \!\!&=&\!\! \frac{1}{|k|}
    \int_{0}^{\infty} dx (1-\cos x)
    w_{+}\! \bigg( \frac{x}{|k|} \bigg)
    \nonumber\\ [4pt]
    &&\!\! - \frac{i}{k}
    \int_{0}^{\infty} dx \sin x
    w_{-}\! \bigg( \frac{x}{|k|} \bigg).
\label{w_k2}
\end{eqnarray}
At first sight, there are two different situations when
$\alpha_{\mathrm{max}} \in [\alpha,1)$ and $\alpha_{\mathrm{max}} \in
[1,2]$. However, because we need to know only the leading term of the
asymptotic expansion of $1-w_{k}$ as $k \to 0$, we can restrict
ourselves to considering Eq.~(\ref{w_k2}) for $\alpha_{ \mathrm{max}}
\in [\alpha,1)$. In this case, using the asymptotic formula
(\ref{w_asymm}) and the standard integrals (\ref{int1}) and
\begin{equation}
    \int_{0}^{\infty} dx \frac{\sin x}{x^{1+\nu}} =
    \frac{\pi}{2\Gamma(1+\nu) \cos(\pi \nu/2)}
\label{int3}
\end{equation}
($0<\nu<1$), one can show that Eq.~(\ref{w_k2}) at $k \to 0$ reduces
to Eq.~(\ref{as_w1}) with the parameters $q$ and $r$ given by the
same Eqs.~(\ref{q}) and (\ref{r}). Since these results hold also for
$\alpha_{ \mathrm{max}} \in [1,2]$, it can be concluded that the
expressions (\ref{as_a3}) and (\ref{limP4}) for the scaling function
$a(t)$ and the limiting probability density $\mathcal{P}(y)$ are
valid not only for $\alpha \in (1,2)$ but also for $\alpha \in
(0,1)$. We note that Eqs.~(\ref{as_a3}) and (\ref{limP4}) at
$\alpha_{+} = \alpha_{-} = \alpha \in (0,1)$ and $u_{+} = u_{-} = u$
are reduced to Eqs.~(\ref{a_heavy}) and (\ref{P_heavy}),
respectively.

\subsection{Jump densities with $\bm{\alpha = 1}$}
\label{3e}

Denoting the first and second terms on the right-hand side of
Eq.~(\ref{w_k2}) by $J_{1}$ and $J_{2}$, respectively, at $\alpha =1$
and $k \to 0$ we obtain
\begin{equation}
    J_{1} \sim \frac{\pi}{2}(u_{+} \delta_{1 \alpha_{+}} +
    u_{-} \delta_{1 \alpha_{-}})|k|
\label{J1}
\end{equation}
and
\begin{eqnarray}
    J_{2} \!\!&\sim&\!\! ik  \int_{c|k|}^{\infty}
    dx \frac{\sin x}{x^{2}} (u_{+}\delta_{1\alpha_{+}}
    - u_{-}\delta_{1\alpha_{-}})
    \nonumber\\ [4pt]
    \!\!&\sim&\!\! i (u_{+}\delta_{1\alpha_{+}}
    - u_{-}\delta_{1\alpha_{-}}) k\ln \frac{1}{|k|},
\label{J2}
\end{eqnarray}
where $c$ is a positive constant. If the parameter
\begin{equation}
    \rho = u_{+}\delta_{1\alpha_{+}} -
    u_{-}\delta_{1\alpha_{-}}
\label{Delta}
\end{equation}
is not equal zero, the term $J_{1}$ can be neglected in comparison
with $J_{2}$. In this case
\begin{equation}
    1 - w_{k} \sim -i\rho k\ln \frac{1}{|k|}
\label{as_w2}
\end{equation}
($k \to 0$) and
\begin{eqnarray}
    \Phi(\kappa) \!\!&=&\!\! -i\rho \kappa
    \lim_{t \to \infty} \frac{a(t)}{V(t)}\ln
    \frac{1}{|\kappa|a(t)}
    \nonumber\\ [4pt]
    \!\!&=&\!\! -i\rho \kappa \lim_{t \to \infty}
    \frac{a(t)}{V(t)}\ln \frac{1}{a(t)}.
\label{Phi4}
\end{eqnarray}

Let us assume that
\begin{equation}
    a(t) \sim \frac{V(t)}{|\rho| \ln [1/V(t)]}
\label{as_a4}
\end{equation}
as $t \to \infty$. Then, taking into account that $V(t) \to 0$ as $t
\to \infty$ and $\lim_{\epsilon \to \infty}\ln\ln \epsilon/ \ln
\epsilon = 0$, one easily finds
\begin{equation}
    \lim_{t \to \infty}\frac{a(t)}{V(t)}
    \ln \frac{1}{a(t)} = \frac{1}{|\rho|}.
\label{relat3}
\end{equation}
Therefore, in this case $\Phi(\kappa) = -i\kappa\,
\mathrm{sgn}(\rho)$ and Eqs.~(\ref{limP2}) yields
\begin{equation}
    \mathcal{P}(y) =  e^{-|y|} H(\rho y).
\label{limP5}
\end{equation}
The comparison of Eqs.~(\ref{limP5}) and (\ref{limP3}) shows that the
limiting probability density at $\alpha = 1$ and $\rho \neq 0$ has
the same form as in the case of jump densities with $l_{1} = \rho$.
Thus, the parameter $\rho$ plays here the role of the first moment of
$w(\xi)$. We note, however, that this analogy is not complete because
at $\alpha = 1$ the first moment $l_{1}$ does not exist. The
difference between the scaling functions (\ref{as_a4}) and
(\ref{as_a2}), which correspond to the limiting densities
(\ref{limP5}) and (\ref{limP3}), has the same origin.

In the opposite case, when $\rho =0$ (this takes place only if
$\alpha_{+} = \alpha_{-} =1$ and $u_{+} = u_{-} = u$), the
contribution of $J_{1}$ becomes dominant and thus
\begin{equation}
    1 - w_{k} \sim  \pi u|k|
\label{as_w3}
\end{equation}
($k \to 0$) and
\begin{equation}
    \Phi(\kappa) = \pi u\kappa \lim_{t \to \infty}
    \frac{a(t)}{V(t)}.
\label{Phi5}
\end{equation}
Choosing the long-time behavior of the scaling function in the form
\begin{equation}
    a(t) \sim \frac{V(t)}{\pi u},
\label{as_a5}
\end{equation}
from Eq.~(\ref{limP2}) with $\Phi(\kappa) = |\kappa|$ we immediately
get
\begin{equation}
    \mathcal{P}(y) =  \frac{1}{\pi} \int_{0}^{\infty}
    d\kappa \frac{\cos(y \kappa)}{1 + \kappa}.
\label{limP6}
\end{equation}
This result is a particular case of Eq.~(\ref{P_heavy}).

\subsection{Jump densities with $\bm{\alpha = 2$}}
\label{3f}

Since $\alpha_{\pm} \leq 2$, the condition $\alpha = 2$ implies that
$\alpha_{+} = \alpha_{-} =2$. It is clear that if $l_{1} \neq 0$,
then the limiting probability density is given by Eq.~(\ref{limP3}).
In contrast, at $l_{1} = 0$ from Eqs.~(\ref{w_k1}) and (\ref{defPhi})
one obtains
\begin{eqnarray}
    1 - w_{k} \!\!&\sim &\!\! (u_{+} + u_{-})k^{2}
    \int_{c|k|}^{\infty} dx \frac{1-\cos x}{x^{3}}
    \nonumber\\ [4pt]
    \!\!&\sim &\!\! \frac{1}{2} (u_{+} + u_{-})
    k^{2} \ln \frac{1}{|k|}
\label{as_w4}
\end{eqnarray}
($k \to 0$, $c$ is a positive parameter) and
\begin{equation}
    \Phi(\kappa) = \frac{1}{2} (u_{+} + u_{-})
    \kappa^{2}\lim_{t \to \infty}\frac{a^{2}(t)}{V(t)}
    \ln \frac{1}{a(t)}.
\label{Phi6}
\end{equation}
If the asymptotic behavior of the scaling function is governed by the
relation
\begin{equation}
    a(t) \sim 2 \sqrt{\frac{V(t)}{(u_{+} + u_{-})
    \ln[1/V(t)]}}
\label{as_a6}
\end{equation}
($t \to \infty$), then
\begin{equation}
    \lim_{t \to \infty} \frac{a^{2}(t)}{V(t)}
    \ln \frac{1}{a(t)} = \frac{2}{u_{+} + u_{-}}.
\label{lim}
\end{equation}
Hence, in this case Eq.~(\ref{Phi6}) reduces to $\Phi(\kappa) =
\kappa^{2}$ and Eq.~(\ref{limP2}) yields
\begin{equation}
    \mathcal{P}(y) =  \frac{1}{\pi} \int_{0}^{\infty}
    d\kappa \frac{\cos(y \kappa)}{1 + \kappa^{2}}
    = \frac{1}{2}\,e^{-|y|}.
\label{limP7}
\end{equation}

The same result was obtained in Ref.~\cite{DYBKL} under the condition
that heavy-tailed jump densities with $\alpha =2$ are symmetric.
However, since the condition $l_{1}=0$ does not imply that $w(-\xi) =
w(\xi)$, the two-sided exponential density (\ref{limP7}) corresponds
to a more wide class jump densities characterized by the conditions
$\alpha =2$ and $l_{1}=0$. It should be noted that the limiting
probability density (\ref{limP7}) as well as the limiting density
(\ref{limP6}) can also be obtained from the general representation
(\ref{limP4}) by taking the limits $\alpha \to 2$ and $\alpha \to 1$,
respectively. But since the scaling functions (\ref{as_a6}) and
(\ref{as_a4}) do not follow from Eq.~(\ref{as_a3}), we considered
these cases separately.

Thus, according to the above analysis, the CTRWs with
superheavy-tailed distributions of waiting time are characterized by
two different classes of limiting probability densities. The first
one is formed by the exponential densities (\ref{limP3}),
(\ref{limP5}), and (\ref{limP7}) that correspond to the jump
densities with (i) $l_{1} \neq 0$ and $|l_{1}|< \infty$, (ii) $\alpha
=1$ and $\rho \neq 0$, and (iii) $\alpha =2$ and $l_{1} = 0$,
respectively. The second one, which describes all other cases, is
constituted by a two-parametric (with parameters $\alpha$ and
$\varphi$) probability density (\ref{limP4}). If $\varphi=0$, then
the limiting probability density (\ref{limP4}) is reduced to the
symmetric one (\ref{P_heavy}), whose properties is well established
\cite{DYBKL}. As for the non-symmetric case, it has never been
studied. However, because of the oscillating character of the
integrand, the use of the limiting density $\mathcal{P}(y)$ in the
form of Eq.~(\ref{limP4}) [we recall that this form of
$\mathcal{P}(y)$ corresponds to $\alpha \in (0,1)$ or $\alpha \in
(1,2)$ and $l_{1} = 0$] is not always convenient. Therefore, to gain
more insight into the analytical properties of $\mathcal{P}(y)$, next
we derive its different representations.

\section{ALTERNATIVE REPRESENTATIONS OF $\bm{\mathcal{P} (y)}$}
\label{rep}

\subsection{Representation of $\bm{\mathcal{P} (y)}$ in terms
of the inverse Mellin transform} \label{Mell}

To derive alternative expressions for the limiting probability
density (\ref{limP4}), we first rewrite it in the form
\begin{equation}
    \mathcal{P}(y) =  \mathcal{P}_{1}(y)  + \mathrm{sgn}(y)
    \mathcal{P}_{2}(y),
\label{limP4b}
\end{equation}
where
\begin{equation}
    \mathcal{P}_{1}(y) = \frac{1}{\pi} \int_{0}^{\infty}
    d\kappa \frac{(1 + \cos \varphi \, \kappa^{\alpha})
    \cos(y\kappa)}{1 + 2\cos\varphi \, \kappa^{\alpha}
    + \kappa^{2\alpha}}
\label{P1}
\end{equation}
and
\begin{equation}
    \mathcal{P}_{2}(y) = \frac{\sin\varphi}{\pi} \int_{0}^{\infty}
    d\kappa \frac{ \kappa^{\alpha} \sin(|y|\kappa)}{1 +
    2\cos\varphi \, \kappa^{\alpha}  + \kappa^{2\alpha}}
\label{P2}
\end{equation}
are even functions of $y$. Then we calculate the Mellin transform of
the functions $\mathcal{P}^{\pm}(y) = \mathcal{P}_{1}(y) \pm
\mathcal{P}_{2}(y)$, which represent $\mathcal{P}(y)$ for positive
and negative $y$: $\mathcal{P}(y)|_{y \gtrless 0} = \mathcal{P}^{
\pm}(y)$. Using the definition of the Mellin transform of a function
$f(y)$, $f_{r} = \mathcal{M} \{ f(y) \} = \int_{0}^{\infty} dy f(y)
y^{r-1}$, and the relation $f_{r} = u_{r}v_{1-r}$ that holds for the
function $f(y) = \int_{0}^{\infty} dx u(yx)v(x)$ \cite{DB}, we obtain
\begin{equation}
    \mathcal{P}^{\pm}_{r} =  \mathcal{M}\{ \cos y\} F_{1-r} \pm
    \mathcal{M}\{ \sin y\} G_{1-r}.
\label{P+-_r}
\end{equation}
Here, according to Ref.~\cite{Erd1}, $\mathcal{M}\{ \cos y\} =
\Gamma(r) \cos (\pi r/2)$ ($0<\mathrm{Re} \,r<1$), $\mathcal{M}\{
\sin y\} = \Gamma(r) \sin (\pi r/2)$ ($-1<\mathrm{Re}\,r<1$),
\begin{eqnarray}
    F_{1-r} \!\!&=&\!\! \frac{1}{\pi} \int_{0}^{\infty} dy \frac{1+
    \cos \varphi\, y^{\alpha}}{1 + 2\cos\varphi\, y^{\alpha} +
    y^{2\alpha}}y^{-r}
    \nonumber\\ [4pt]
    \!\!&=&\!\! \frac{1}{\pi \alpha} \int_{0}^{\infty} dy \frac{1+
    \cos \varphi\, y}{1 + 2\cos\varphi\, y + y^{2}}
    y^{\frac{1-r}{\alpha} - 1}
    \nonumber\\ [4pt]
    \!\!&=&\!\! \frac{\cos[\varphi (1-r)/\alpha]}
    {\alpha \sin[\pi (1-r)/\alpha]}
\label{F_1-r}
\end{eqnarray}
($1-\alpha<\mathrm{Re}\,r<1$), and
\begin{eqnarray}
    G_{1-r} \!\!&=&\!\! \frac{\sin \varphi}{\pi} \int_{0}^{\infty}
    dy \frac{ y^{\alpha-r}}{1 + 2\cos\varphi\, y^{\alpha} +
    y^{2\alpha}}
    \nonumber\\ [4pt]
    \!\!&=&\!\! \frac{\sin \varphi}{\pi \alpha} \int_{0}^{\infty}
    dy \frac{y^{\frac{1-r}{\alpha}}}{1 + 2\cos\varphi\, y + y^{2}}
    \nonumber\\ [4pt]
    \!\!&=&\!\! \frac{\sin[\varphi (1-r)/\alpha]}
    {\alpha \sin[\pi (1-r)/\alpha]}
\label{G_1-r}
\end{eqnarray}
($1-\alpha<\mathrm{Re}\,r<1 + \alpha$).

Collecting the above results, from Eq.~(\ref{P+-_r}) one finds
\begin{equation}
    \mathcal{P}^{\pm}_{r} = \frac{\Gamma(r) \sin[
    (\pi \alpha/2 \pm \varphi)(1 - r)/\alpha]}
    {\alpha\sin[\pi(1-r)/\alpha]}
\label{P+-_r2}
\end{equation}
with $\max(1-\alpha, 0)<\mathrm{Re}\,r<1$. From this, using the
definition of the inverse Mellin transform, $\mathcal{M}^{-1}
\{f_{r}\} = f(y)= (2\pi i)^{-1} \int_{c-i\infty}^{c+i\infty} dr
f_{r}y^{-r}$, the relation $\mathcal{P}(y) |_{y \gtrless 0} =
\mathcal{P}^{ \pm}(y)$ and Eq.~(\ref{P+-_r2}), we obtain the limiting
probability density (\ref{limP4}) in terms of the inverse Mellin
transform
\begin{equation}
    \mathcal{P}(y) = \frac{1}{2\pi i} \int_{c-i\infty}
    ^{c+i\infty} dr \frac{\Gamma(r)\sin\! \big[\phi(y)
    \frac{1- r}{\alpha}\big]}{\alpha\sin\! \big(\pi
    \frac{1- r}{\alpha}\big)} |y|^{-r},
\label{limP_M}
\end{equation}
where $\max(1-\alpha, 0)<c<1$  and
\begin{equation}
    \phi(y) = \frac{\pi \alpha}{2} +
    \mathrm{sgn}(y) \varphi.
\label{psi}
\end{equation}

It is worth to note that both representations of $\mathcal{P}(y)$,
(\ref{limP4}) and (\ref{limP_M}), are valid for all values of the
lowest tail index $\alpha$ from the interval $(0,2]$. But since the
limiting probability densities at $\alpha=1$ and $\alpha=2$ have
already been determined in Secs.~\ref{3e} and \ref{3f}, further we
examine Eq.~(\ref{limP_M}) for $\alpha \in (0,1)$ and $\alpha \in
(1,2)$ only. There are four different cases associated with these
intervals, which we consider separately below.

\subsubsection{$\alpha \in (0,1)$, $\alpha_{+} \neq \alpha_{-}$}

In this case, Eq.~(\ref{def_phi}) together with Eqs.~(\ref{q}),
(\ref{r}) and (\ref{relat1}) yields $\cos \varphi = \cos(\pi
\alpha/2)$ and $\sin \varphi = (\delta_{\alpha \alpha_{+}} -
\delta_{\alpha \alpha_{-}}) \sin(\pi \alpha/2)$. From the last two
equations it follows that $\varphi = (\delta_{\alpha \alpha_{+}} -
\delta_{\alpha \alpha_{-}}) \pi\alpha/2$, and Eq.~(\ref{psi}) reduces
to
\begin{equation}
    \phi(y) = [1 + \mathrm{sgn}(\sigma y)]
    \frac{\pi \alpha}{2},
\label{psi1}
\end{equation}
where $\sigma = \delta_{\alpha \alpha_{+}} - \delta_{\alpha
\alpha_{-}}\! = \mathrm{sgn}(\alpha_{-} - \alpha_{+})$. Denoting
$\phi_{\mathrm{sgn}(\sigma y)} = \phi(y)$, Eq.~(\ref{psi1}) yields
$\phi_{+}=\pi \alpha$ and $\phi_{-}=0$. The last condition means that
$\mathcal{P}(y) =0$ as $\sigma y<0$. In other words, in the case when
$\alpha \in (0,1)$ and $\alpha_{+} \neq \alpha_{-}$ the limiting
probability density $\mathcal{P}(y)$ is one-sided. According to
Eq.~(\ref{limP_M}), it can be represented as
\begin{equation}
    \mathcal{P}(y) = \frac{H(\sigma y)}
    {2\pi i} \int_{c-i\infty}^{c+i\infty} dr
    \frac{\Gamma(r) \sin\! \big(\phi_{+} \frac{1-
    r}{\alpha}\big)} {\alpha \sin\!
    \big(\pi\frac{1- r}{\alpha}\big)} |y|^{-r}
\label{limP_M1}
\end{equation}
with $\phi_{+}= \pi\alpha$.

\subsubsection{$\alpha \in (1,2)$, $\alpha_{+} \neq \alpha_{-}$,
$l_{1}=0$}

For these conditions, Eq.~(\ref{def_phi}) leads to the equations
$\cos \varphi = -\cos(\pi \alpha/2)$ and $\sin \varphi = - \sigma
\sin(\pi \alpha/2)$, whose solution is given by $\varphi = -\sigma
(\pi - \pi\alpha/2)$. In this case Eq.~(\ref{psi1}) reads
\begin{equation}
    \phi(y) = \frac{\pi \alpha}{2} -
    \mathrm{sgn}(\sigma y)\Big(\pi -
    \frac{\pi \alpha}{2} \Big),
\label{psi2}
\end{equation}
and so $\phi_{+}= \pi(\alpha-1)$ and $\phi_{-}= \pi$. Therefore,
using the inverse Mellin transform of the $\Gamma$ function
\cite{Erd1}
\begin{equation}
    \frac{1}{2\pi i} \int_{c-i\infty}^{c+i\infty}
    dr \Gamma(r)|y|^{-r} = e^{-|y|}
\label{relat4}
\end{equation}
($c>0$), Eq.~(\ref{limP_M}) can be rewritten as
\begin{eqnarray}
    \mathcal{P}(y) \!\!&=&\!\! \frac{H(\sigma y)}
    {2\pi i } \int_{c-i\infty}^{c+i\infty} dr
    \frac{\Gamma(r) \sin\! \big(\phi_{+} \frac{1-
    r}{\alpha}\big)} {\alpha\sin\!
    \big(\pi\frac{1- r}{\alpha}\big)}|y|^{-r}
    \nonumber\\ [4pt]
    &&\!\! +\, H(-\sigma y) \frac{1}{\alpha}\, e^{-|y|},
\label{limP_M2}
\end{eqnarray}
where $\phi_{+}= \pi(\alpha-1)$.

Thus, if $\alpha \in (1,2)$, $\alpha_{+} \neq \alpha_{-}$ and
$l_{1}=0$, then, in contrast to the previous case, the limiting
probability density is two-sided. As is clear from
Eq.~(\ref{limP_M2}), this density exhibits an exponential decay at
$y>0$ if $\alpha =\alpha_{-}$ or at $y<0$ if $\alpha =\alpha_{+}$. We
note also that, to avoid the double contribution of the point $y=0$,
the condition $H(y\sigma)|_{y=0} = H(\sigma)$ is assumed to hold.

\subsubsection{$\alpha_{+} = \alpha_{-} = \alpha \in (0,1)$,
$u_{+} \neq u_{-}$}

Under these conditions, Eq.~(\ref{def_phi}) can be expressed as
\begin{equation}
    \begin{array}{ll}
    \displaystyle \cos\varphi = \frac{(u_{+} + u_{-})
    \cos(\pi\alpha/2)}{\sqrt{u_{+}^{2}
    + u_{-}^{2} + 2\cos(\pi\alpha) u_{+}u_{-}}},
    \\ [18pt]
    \displaystyle \sin\varphi = \frac{(u_{+} - u_{-})
    \sin(\pi\alpha/2)}{\sqrt{u_{+}^{2}
    + u_{-}^{2} + 2\cos(\pi\alpha) u_{+}u_{-}}}.
    \end{array}
\label{cs3}
\end{equation}
Introducing the notation
\begin{equation}
    \epsilon = \frac{u_{+} - u_{-}}{u_{+} + u_{-}},
\label{eps}
\end{equation}
from Eq.~(\ref{cs3}) one obtains
\begin{equation}
    \varphi = \mathrm{sgn}(\epsilon)\arctan\!\Big[
    |\epsilon| \tan\! \left(\frac{\pi \alpha}{2}
    \right)\! \Big],
\label{phi3}
\end{equation}
where $\arctan(x)$ denotes the principal value of the inverse tangent
function. Finally, representing the two-valued function (\ref{psi})
as $\phi(y) = \phi_{\mathrm{sgn}(\epsilon y)}$, where
\begin{equation}
    \phi_{\pm} = \frac{\pi \alpha}{2} \pm \arctan\!
    \Big[|\epsilon| \tan\! \left(\frac{\pi \alpha}{2}
    \right)\! \Big],
\label{psi3}
\end{equation}
and using Eq.~(\ref{limP_M}), we find the following two-sided
limiting probability density:
\begin{eqnarray}
    \mathcal{P}(y) \!\!&=&\!\! \frac{H(\epsilon y)}
    {2\pi i } \int_{c-i\infty}^{c+i\infty}
    dr \frac{\Gamma(r) \sin\!\big(\phi_{+}
    \frac{1 - r}{\alpha}\big)}{\alpha\sin\!
    \big(\pi\frac{1- r}{\alpha}\big)}|y|^{-r}
    \nonumber\\ [4pt]
    &&\!\! +  \frac{H(-\epsilon y)}
    {2\pi i } \int_{c-i\infty}^{c+i\infty}dr
    \frac{\Gamma(r) \sin\!\big(\phi_{-}
    \frac{1 - r}{\alpha}\big)}{\alpha\sin\!
    \big(\pi\frac{1- r}{\alpha}\big)}|y|^{-r}.
    \nonumber\\
\label{limP_M3}
\end{eqnarray}
Since $\alpha \in (0,1)$ and $|\epsilon|<1$, one can easily check
that $\arctan [|\epsilon| \tan (\pi \alpha/2)] \in (0, \pi
\alpha/2)$, and so $\pi \alpha/2<\phi_{+}<\pi \alpha$,
$0<\phi_{-}<\pi \alpha/2$, and $\phi_{+}>\phi_{-}$.

\subsubsection{$\alpha_{+} = \alpha_{-} = \alpha \in (1,2)$,
$u_{+} \neq u_{-}$, $l_{1}=0$}

In this last case, the limiting probability density is given by the
same formula (\ref{limP_M3}). To find the parameters $\phi_{+}$ and
$\phi_{-}$, we first write equations
\begin{equation}
    \begin{array}{ll}
    \displaystyle \cos\varphi = -\frac{(u_{+} + u_{-})
    \cos(\pi\alpha/2)}{\sqrt{u_{+}^{2}
    + u_{-}^{2} + 2\cos(\pi\alpha) u_{+}u_{-}}},
    \\ [18pt]
    \displaystyle \sin\varphi = -\frac{(u_{+} - u_{-})
    \sin(\pi\alpha/2)}{\sqrt{u_{+}^{2}
    + u_{-}^{2} + 2\cos(\pi\alpha) u_{+}u_{-}}},
    \end{array}
\label{cs2}
\end{equation}
which follow from Eq.~(\ref{def_phi}). Since their solution can be
represented in the same form as Eq.~(\ref{phi3}), the parameters
$\phi_{+}$ and $\phi_{-}$ can also be determined from
Eq.~(\ref{psi3}). However, because $\arctan [|\epsilon| \tan (\pi
\alpha/2)] \in (\pi \alpha/2 - \pi, 0)$, in contrast to the previous
case we have $\pi(\alpha -1)<\phi_{+}<\pi \alpha/2$, $\pi
\alpha/2<\phi_{-}<\pi$, and $\phi_{+}<\phi_{-}$.

\subsection{Representation of $\bm{\mathcal{P} (y)}$
in terms of the Laplace transform} \label{Lap}

To derive the limiting probability density in terms of the Laplace
transform, in Eq.~(\ref{limP_M}) we first introduce a new variable of
integration $\eta = (1-r)/\alpha$ and use the integral representation
$\Gamma(r) = \int_{0}^{\infty} dz e^{-z}z^{r-1}$ ($\mathrm{Re}\,
r>0$) of the $\Gamma$ function \cite{Erd2}. This yields
\begin{eqnarray}
    \mathcal{P}(y) \!\!&=&\!\! \frac{1}{2\pi i }
    \int_{c-i\infty}^{c+i\infty} d\eta\, \Gamma(1-
    \alpha\eta) \frac{\sin[\phi(y)\eta]}{\sin(\pi\eta)}
    |y|^{\alpha\eta -1}
    \nonumber\\ [4pt]
    \!\!&=&\!\!  \frac{1}{2\pi i }\int_{0}^{\infty}
    dz e^{-z} \int_{c-i\infty}^{c+i\infty}d\eta
    \frac{\sin[\phi(y)\eta]}{|y|\sin(\pi\eta)}\!
    \left(\frac{z^{\alpha}}{|y|^{\alpha}}
    \right)^{\!-\eta}
    \nonumber\\
\label{limP_tr}
\end{eqnarray}
with $0<c<\min\{1,1/\alpha\}$. Then, taking into account the relation
\cite{PK}
\begin{equation}
    \frac{1}{2 i} \int_{c-i\infty}^{c+i\infty}d\eta
    \frac{\sin(\vartheta\eta)}{\sin(\pi\eta)} z^{-\eta} =
    \frac{\sin\vartheta\, z}{1+2\cos\vartheta\, z + z^{2}}
\label{relat5}
\end{equation}
($-\pi<\vartheta<\pi$) and changing in Eq.~(\ref{limP_tr}) the
integration variable from $z$ to $x= z/|y|$, we obtain the desired
representation of the limiting probability density
\begin{equation}
    \mathcal{P}(y) = \frac{1}{\pi} \int_{0}^{\infty}
    dx\, e^{-|y|x}\frac{\sin[\phi(y)] x^{\alpha}}
    {1 + 2\cos[\phi(y)] x^{\alpha} + x^{2\alpha}}.
\label{limP_L}
\end{equation}
Note that, although the relation (\ref{relat5}) is not valid for
$\vartheta = \pi$, Eq.~(\ref{limP_L}) at $\phi(y) = \pi$ gives a
correct result if $\mathcal{P}(y)|_{\phi(y) = \pi}$ is interpreted as
the limit $\lim_{\zeta \to 0} \mathcal{P}(y)|_{\phi(y) = \pi -
\zeta}$.

The limiting probability density $\mathcal{P}(y)$ in the form of
Eq.~(\ref{limP_L}) is useful to establish its general properties. In
particular, directly from this representation it follows that
$\mathcal{P}(y)\geq 0$ [i.e., $\mathcal{P}(y)$ is in fact the
probability density], $d\mathcal{P}(y)/d|y|\leq 0$ ($y \neq 0$), and
$\max{\mathcal{P}(y)} = \mathcal{P}(0)$. Moreover, because of the
exponential factor in the integrand, the representation
(\ref{limP_L}) is the most suitable for the numerical evaluation of
$\mathcal{P}(y)$ at large $|y|$.

For convenience of use, we write below the representations of
$\mathcal{P} (y)$ in terms of the Laplace transform for all four
cases considered in Sec.~\ref{Mell}.

\subsubsection{$\alpha \in (0,1)$, $\alpha_{+} \neq \alpha_{-}$}

According to Eqs.~(\ref{limP_M1}) and (\ref{limP_L}), in this case
\begin{equation}
    \mathcal{P}(y) = \frac{H(\sigma y)}{\pi}
    \int_{0}^{\infty}dx\, e^{-|y|x}\frac{
    \sin(\phi_{+}) x^{\alpha}} {1 +
    2\cos(\phi_{+})x^{\alpha} + x^{2\alpha}}
\label{limP_L1}
\end{equation}
with $\phi_{+} = \pi \alpha$. Since at $y=0$ the above integral
diverges, one gets $\mathcal{P} (y)|_{\sigma y \to +0} =\infty$.
Then, using the standard integral \cite{PBM}
\begin{equation}
    \int_{0}^{\infty}dz\frac{z^{\nu-1}}{1 + 2\cos\vartheta\,
    z + z^{2}} = \frac{\pi \sin[\vartheta(1-\nu)]}
    {\sin \vartheta \sin(\pi\nu)}
\label{relat6}
\end{equation}
($0<|\vartheta|<\pi, 0<\nu<2$) and taking into account that at $\nu =
1$ the right-hand side of Eq.~(\ref{relat6}) equals
$\vartheta/\sin\vartheta$, one can easily make sure that the
normalization condition for $\mathcal{P}(y)$ holds:
\begin{eqnarray}
    \int_{-\infty}^{\infty} dy \mathcal{P}(y)
    \!\!&=&\!\! \frac{1}{\pi} \int_{0}^{\infty} dx
    \frac{\sin(\pi\alpha) x^{\alpha-1}}
    {1 + 2\cos(\pi\alpha) x^{\alpha} + x^{2\alpha}}
    \nonumber\\ [4pt]
    \!\!&=&\!\!  \frac{1}{\pi\alpha} \int_{0}^{\infty}
    dz\frac{\sin(\pi\alpha)} {1 + 2\cos(\pi\alpha)z
    + z^{2}} =1.
    \nonumber\\
\label{norm1}
\end{eqnarray}

The main feature of the limiting probability density in the reference
case is that it is one-sided with $\mathcal{P}(y) = 0$ at $y>0$ if
$\alpha_{+}> \alpha_{-}$ or at $y<0$ if $\alpha_{-}> \alpha_{+}$.
These conditions show that $\mathcal{P}(y)$ is concentrated on the
semi-axis where the tail index is the smallest, i.e., where the
probability of long-distance jumps of a particle is the largest. For
clarity, we note that the total probability of jumps in this
direction, $W_{\mathrm{sgn}(\sigma)} = \int_{0}^{ \infty} d\xi
w[\mathrm{sgn} (\sigma) \xi]$, can be even less than the total
probability $W_{-\mathrm{sgn} (\sigma)} = 1 - W_{\mathrm{sgn}
(\sigma)}$ of jumps in the opposite direction.

\begin{figure}
    \centering
    \includegraphics[totalheight=5cm]{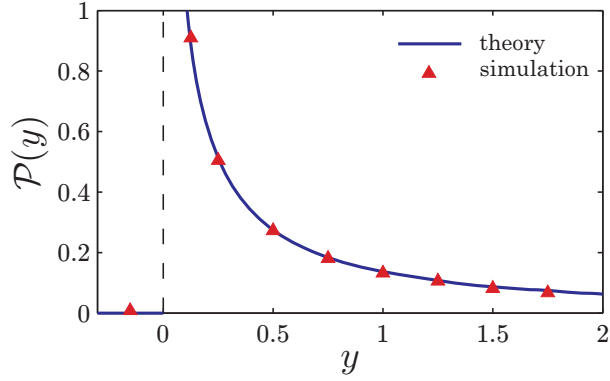}
    \caption{\label{fig1} (Color online) The limiting
    probability density at $\alpha_{+} = \alpha =1/2$
    and $\alpha_{-}> \alpha_{+}$.
    The solid blue line represents the theoretical result
    (\ref{limP_L1}) and the simulation results (see
    Sec.~\ref{Num}) are indicated by red triangles.}
\end{figure}

The behavior of the limiting probability density at $\alpha \in
(0,1)$ and $\alpha_{+} \neq \alpha_{-}$ is illustrated in
Fig.~\ref{fig1}.

\subsubsection{$\alpha \in (1,2)$, $\alpha_{+} \neq \alpha_{-}$,
$l_{1}=0$}

For these conditions, the limiting probability density in terms of
the Laplace transform reads
\begin{eqnarray}
    \mathcal{P}(y) \!\!&=&\!\! \frac{H(\sigma y)}{\pi}
    \int_{0}^{\infty}dx\, e^{-|y|x}\frac{\sin(\phi_{+})
    \, x^{\alpha}} {1 + 2\cos(\phi_{+})\, x^{\alpha} +
    x^{2\alpha}}
    \nonumber\\ [4pt]
    &&\!\! +\, H(-\sigma y) \frac{1}{\alpha}\, e^{-|y|},
\label{limP_L2}
\end{eqnarray}
where $\phi_{+} = \pi(\alpha - 1)$. It can be shown with the help of
Eq.~(\ref{relat6}) that $\mathcal{P}(y)$ is normalized and
\begin{eqnarray}
    \mathcal{P}(y)|_{\sigma y \to +0}
    \!\!&=&\!\! \frac{1}{\pi} \int_{0}^{\infty} dx
    \frac{\sin(\phi_{+}) x^{\alpha}}
    {1 + 2\cos(\phi_{+}) x^{\alpha} + x^{2\alpha}}
    \nonumber\\ [4pt]
    \!\!&=&\!\!  \frac{1}{\pi\alpha} \int_{0}^{\infty}
    dz\frac{\sin(\phi_{+}) z^{1/\alpha}} {1 +
    2\cos(\phi_{+})z + z^{2}}
    \nonumber\\ [4pt]
    \!\!&=&\!\! \frac{\sin(\phi_{+}/\alpha)}{\alpha
    \sin(\pi/\alpha)}.
\label{P|y=+0}
\end{eqnarray}
Since $\phi_{+} = \pi(\alpha - 1)$ and, in accordance with
Eq.~(\ref{limP_L2}), $\mathcal{P}(y)|_{\sigma y \to -0} = 1/\alpha$,
we obtain
\begin{equation}
    \mathcal{P}(0) = \mathcal{P}(y)|_{\sigma y
    \to \pm 0} = \frac{1}{\alpha}.
\label{P(0)1}
\end{equation}

Thus, in contrast to the previous case, the limiting probability
density is two-sided and is bounded at the origin. One branch of
$\mathcal{P}(y)$, left if $\alpha_{-} >\alpha_{+}$ or right if
$\alpha_{+} >\alpha_{-}$, is purely exponential, and the other is
heavy-tailed (see also Sec.~\ref{beh}). As before, the latter is
concentrated on the semi-axis where the tail index is the smallest.
Interestingly, the probability $\int_{0}^{\infty} dy \mathcal{P}
[-\mathrm{sgn}(\sigma) y] = 1/\alpha$ that $\sigma Y(\infty)<0$,
i.e., the total probability defined by the exponential branch, is
larger than $1/2$.

\begin{figure}
    \centering
    \includegraphics[totalheight=5cm]{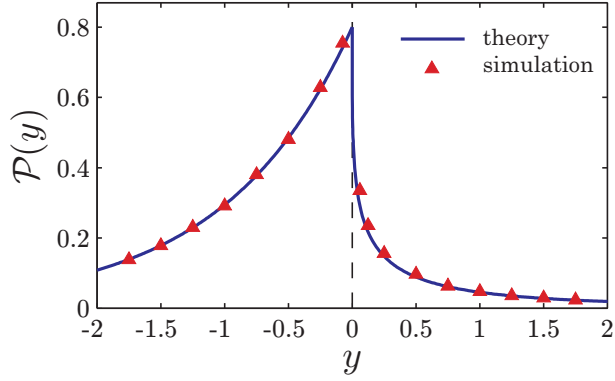}
    \caption{\label{fig2} (Color online) The limiting
    probability density at $\alpha_{+} = \alpha =5/4$
    and $\alpha_{-}> \alpha_{+}$.
    The solid line with an exponential left branch is
    obtained from Eq.~(\ref{limP_L2}) and triangles
    show the simulation results.}
\end{figure}

Figure \ref{fig2} illustrates the behavior of $\mathcal{P}(y)$ in
this case.

\subsubsection{$\alpha_{+} = \alpha_{-} = \alpha \in (0,1)$,
$u_{+} \neq u_{-}$}

From Eqs.~(\ref{limP_M3}) and (\ref{limP_L}) it follows that
\begin{eqnarray}
    \mathcal{P}(y) \!\!&=&\!\! \frac{H(\epsilon y)}
    {\pi} \int_{0}^{\infty} dx\, e^{-|y|x}\frac{
    \sin(\phi_{+}) x^{\alpha}}
    {1 + 2\cos(\phi_{+}) x^{\alpha} + x^{2\alpha}}
    \nonumber\\ [4pt]
    &&\!\! +  \frac{H(-\epsilon y)}
    {\pi} \int_{0}^{\infty}
    dx\, e^{-|y|x}\frac{\sin(\phi_{-}) x^{\alpha}}
    {1 + 2\cos(\phi_{-}) x^{\alpha} + x^{2\alpha}},
    \nonumber\\
\label{limP_L3}
\end{eqnarray}
where the parameters $\phi_{+}$ and $\phi_{-}$ are given by
Eq.~(\ref{psi3}). Since $\alpha<1$, one has $\mathcal{P}(0) = \infty$
and, using again Eq.~(\ref{relat6}), it can be verified that
$\mathcal{P}(y)$ is normalized.

The comparison with the first case shows that, while the difference
in the tail indexes $\alpha_{+}$ and $\alpha_{-}$ leads to a strongly
asymmetric one-sided $\mathcal{P}(y)$, the difference in the
parameters $u_{+}$ and $u_{-}$ (under condition that $\alpha_{+} =
\alpha_{-}$) results in a less asymmetric two-sided $\mathcal{P}(y)$.
According to Eq.~(\ref{limP_L3}), both branches of the limiting
probability density have heavy tails characterized by the same tail
index $\alpha$ (see also Sec.~\ref{beh}).

\begin{figure}
    \centering
    \includegraphics[totalheight=5cm]{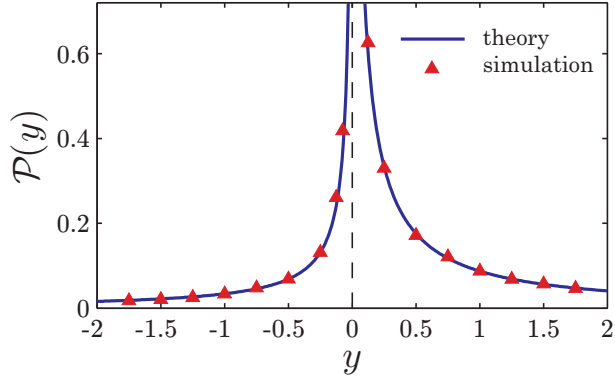}
    \caption{\label{fig3} (Color online) The limiting
    probability density at $\alpha_{+} = \alpha_{-} =
    \alpha = 1/2$ and $\epsilon = 1/3$.
    The solid line and triangles represent the limiting
    probability density (\ref{limP_L3}) and simulation
    data, respectively.}
\end{figure}

In this case, the behavior of $\mathcal{P}(y)$ is illustrated in
Fig.~\ref{fig3}.

\subsubsection{$\alpha_{+} = \alpha_{-} = \alpha \in (1,2)$,
$u_{+} \neq u_{-}$, $l_{1}=0$}

As in the previous section, the limiting probability density and the
parameters $\phi_{+}$ and $\phi_{-}$ are determined by
Eqs.~(\ref{limP_L3}) and (\ref{psi3}), respectively. The striking
difference between the behavior of $\mathcal{P}(y)$ in these cases is
that now $\mathcal{P}(0)<\infty$. To find $\mathcal{P}(0)$, we use
Eq.~(\ref{limP_L3}), which together with the standard integral
(\ref{relat6}) yields
\begin{equation}
    \mathcal{P}(y)|_{\epsilon y \to \pm 0} =
    \frac{\sin(\phi_{\pm}/\alpha)}
    {\alpha \sin(\pi/\alpha)}.
\label{P|y=+-0}
\end{equation}
Using Eq.~(\ref{psi3}), one obtains $\sin(\phi_{+}/\alpha) =
\sin(\phi_{-}/\alpha)$, and so $\mathcal{P}(y)|_{\epsilon y \to \pm
0} = \mathcal{P}(0)$, where
\begin{equation}
    \mathcal{P}(0) = \frac{1}{\alpha \sin(\pi/\alpha)}
    \cos\! \left\{ \frac{1}{\alpha}\arctan\!\Big[
    |\epsilon| \tan\! \left(\frac{\pi\alpha}{2}
    \right)\! \Big]\! \right\}.
\label{P(0)2}
\end{equation}
Comparing with the second case, we again observe that the difference
in $\alpha_{+}$ and $\alpha_{-}$ causes more change in the behavior
of branches of the limiting probability density than the difference
in $u_{+}$ and $u_{-}$ at $\alpha_{+} = \alpha_{-}$.

\begin{figure}
    \centering
    \includegraphics[totalheight=5cm]{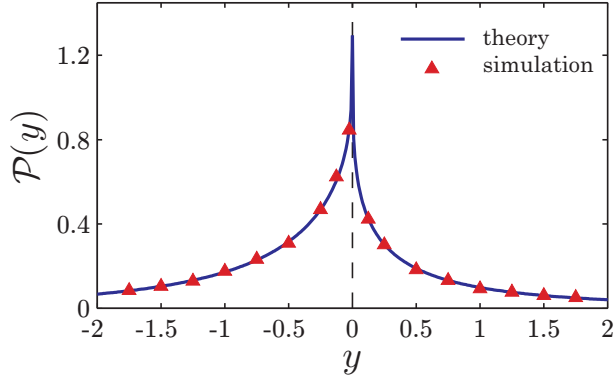}
    \caption{\label{fig4} (Color online) The limiting
    probability density at $\alpha_{+} = \alpha_{-} =
    \alpha = 5/4$ and $\epsilon = (\sqrt{2}-1)/
    (\sqrt{2}+1)$.
    As in the previous case, the solid line is
    obtained from Eq.~(\ref{limP_L3}) and triangles
    indicate the simulation results.}
\end{figure}

The behavior of $\mathcal{P}(y)$ in this last case is shown in
Fig.~\ref{fig4}.

\subsection{Representation of $\bm{\mathcal{P} (y)}$
in terms of the Fox $\bm{H}$ function}

Since the Fox $H$ function is one of the most general special
functions and many of its properties are well studied, it is also
reasonable to express the limiting probability density in terms of
this function. For this purpose, we first use the reflection formula
\cite{Erd2} $\Gamma(z) \Gamma(1-z) = \pi/\sin(\pi z)$ to obtain
\begin{equation}
    \frac{\sin\! \big[\phi(y)\frac{1- r}{\alpha}\big]}
    {\sin\! \big(\pi\frac{1- r}{\alpha}\big)} =
    \frac{\Gamma \big(\frac{1- r}{\alpha}\big)
    \Gamma \big(1 - \frac{1- r}{\alpha}\big)}
    {\Gamma \big[\phi(y)\frac{1- r}{\pi\alpha}\big]
    \Gamma \big[1-\phi(y)\frac{1- r}{\pi\alpha}\big]}.
\label{relat7}
\end{equation}
Then, substituting this relation into Eq.~(\ref{limP_M}), one gets
\begin{equation}
    \mathcal{P}(y) = \frac{1}{2\pi i \alpha}
    \int_{c-i\infty}^{c+i\infty}\! dr \frac{\Gamma(r)
    \Gamma \big(\frac{1- r}{\alpha}\big)
    \Gamma \big(1 - \frac{1- r}{\alpha}\big)}
    {\Gamma \big[\phi(y)\frac{1- r}{\pi\alpha}\big]
    \Gamma \big[1-\phi(y)\frac{1- r}{\pi\alpha}\big]}
    |y|^{-r}.
\label{limP_Mb}
\end{equation}

On the other hand, the $H$ function can be defined in the form of a
Mellin-Barnes integral as follows \cite{MSH}:
\begin{eqnarray}
    H_{p,q}^{m,n}\! \left[ y \Big |
    \begin{array}{lcl}
        (a_{p},A_{p})\\
        (b_{q},B_{q})
    \end{array}
    \right]\!
    \!\!&=&\!\! H_{p,q}^{m,n}\! \left[ y \Big |
    \begin{array}{lcl}
        (a_{1},A_{1}),\ldots,(a_{p},A_{p})\\
        (b_{1},B_{1}),\ldots,(b_{q},B_{q})
    \end{array}
    \right]
    \nonumber\\ [4pt]
    \!\!&=&\!\! \frac{1}{2\pi i} \int_{L} dr
    \Theta_{r} y^{-r},
\label{defH}
\end{eqnarray}
where
\begin{equation}
    \Theta_{r} =\frac{\prod_{j=1}^{m}\Gamma(b_{j} + B_{j}r)
    \prod_{j=1}^{n}\Gamma(1 - a_{j} - A_{j}r)}
    {\prod_{j=m+1}^{q}\Gamma(1 - b_{j} - B_{j}r)
    \prod_{j=n+1}^{p}\Gamma(a_{j} + A_{j}r)},
\label{Theta}
\end{equation}
$m,n,p,q$ are whole numbers, $0 \leq m \leq q$, $0 \leq n \leq p$,
$a_{j}$ and $b_{j}$ are real or complex numbers, $A_{j}, B_{j} >0$,
$L$ is a suitable contour in the complex $r$-plane which separates
the poles of the $\Gamma$ functions $\Gamma(b_{j} + B_{j}r)$ from the
poles of the $\Gamma$ functions $\Gamma(1 - a_{j} - A_{j}r)$, and the
empty product is assumed to be equal to 1. Therefore, comparing
Eq.~(\ref{Theta}) with the integrand in Eq.~(\ref{limP_Mb}), we
obtain the following representation of the limiting probability
density through the $H$ function:
\begin{equation}
    \mathcal{P}(y) = \frac{1}{\alpha}H_{2,3}^{2,1}
    \Bigg[ |y| \bigg |
        \begin{array}{lcl}
        (1-\frac{1}{\alpha},\frac{1}{\alpha}),
        (1-\frac{\phi(y)}{\pi\alpha},\frac{\phi(y)}{\pi\alpha})
        \\[2pt]
        (0,1),(1-\frac{1}{\alpha},\frac{1}{\alpha}),
        (1-\frac{\phi(y)}{\pi\alpha},\frac{\phi(y)}{\pi\alpha})
        \end{array}
    \!\Bigg].
\label{limP_H}
\end{equation}

Using this general formula, it is not difficult to find the
corresponding representations for all four cases considered above.
But here we focus on the first one for which Eq.~(\ref{limP_H}) can
be further simplified. Indeed, since in this case $\phi(y)
=\phi_{\mathrm{sgn}(\sigma y)}$ with $\phi_{+}=\pi \alpha$ and
$\phi_{-}=0$, from Eq.~(\ref{limP_H}) and the reduction formula
\cite{MSH}
\begin{eqnarray}
    &H_{p,q}^{m,n}\! \left[ y \Big |
    \begin{array}{lcl}
        (a_{1},A_{1}),\ldots,(a_{p-1},A_{p-1}),
        (b_{1},B_{1})\\
        (b_{1},B_{1}),\ldots,(b_{q},B_{q})
    \end{array}
    \right]&
    \nonumber\\ [4pt]
    &= H_{p-1,q-1}^{m-1,n}\! \left[ y \Big |
    \begin{array}{lcl}
        (a_{1},A_{1}),\ldots,(a_{p-1},A_{p-1})\\
        (b_{2},B_{2}),\ldots,(b_{q},B_{q})
    \end{array}
    \right]&
\label{relat8}
\end{eqnarray}
($m\geq 1, p>n$) we obtain
\begin{eqnarray}
    \mathcal{P}(y) \!\!&=&\!\! \frac{H(\sigma y)}
    {\alpha}H_{2,3}^{2,1}
    \Bigg[ |y| \bigg |
        \begin{array}{lcl}
        (1-\frac{1}{\alpha},\frac{1}{\alpha}),(0,1)
        \\[2pt]
        (0,1),(1-\frac{1}{\alpha},\frac{1}{\alpha}),(0,1)
        \end{array}
    \!\Bigg]
    \nonumber\\ [4pt]
    \!\!&=&\!\! \frac{H(\sigma y)}
    {\alpha}H_{1,2}^{1,1}
    \Bigg[ |y| \bigg |
        \begin{array}{lcl}
        (1-\frac{1}{\alpha},\frac{1}{\alpha})
        \\[2pt]
        (1-\frac{1}{\alpha},\frac{1}{\alpha}),(0,1)
        \end{array}
    \!\Bigg].
\label{limP_H1}
\end{eqnarray}
Then, using the relation
\begin{equation}
    H_{p,q}^{m,n}\! \left[ y \Big |
    \begin{array}{lcl}
        (a_{p},A_{p})\\
        (b_{q},B_{q})
    \end{array}
    \right]\! = \frac{\chi}{y^{\lambda \chi}}
    H_{p,q}^{m,n}\! \left[ y^{\chi} \Big |
    \begin{array}{lcl}
        (a_{p} + \lambda \chi A_{p},\chi A_{p})\\
        (b_{q} + \lambda \chi B_{q},\chi B_{q})
    \end{array}
    \right]
\label{relat9}
\end{equation}
($\chi>0, -\infty<\lambda<\infty$) with $\chi = \alpha$ and $\lambda
= 1/\alpha -1$, Eq.~(\ref{limP_H1}) can be reduced to
\begin{equation}
    \mathcal{P}(y) = \frac{H(\sigma y)}{|y|^{1-\alpha}}
    \, H_{1,2}^{1,1}\! \left[ |y|^{\alpha} \Big |
    \begin{array}{lcl}
        (0,1)\\
        (0,1), (1-\alpha, \alpha)
    \end{array}
    \right]\! .
\label{limP_H2}
\end{equation}
Finally, since this $H$ function  is closely related to the
generalized Mittag-Leffler function $E_{\alpha,\beta}(z)$ \cite{MSH},
\begin{equation}
    H_{1,2}^{1,1}\! \left[ -z \Big |
    \begin{array}{lcl}
        (0,1)\\
        (0,1), (1-\beta, \alpha)
    \end{array}
    \right]\! = E_{\alpha,\beta}(z)
\label{defM-L}
\end{equation}
($\alpha, \beta\, >0$), for the limiting probability density in the
considered case, when $\alpha \in (0,1)$ and $\alpha_{+} \neq
\alpha_{-}$, one gets
\begin{equation}
    \mathcal{P}(y) = \frac{H(\sigma y)}{|y|^{1-\alpha}}
    \, E_{\alpha,\alpha}(-|y|^{\alpha}).
\label{limP_M-L}
\end{equation}

The usefulness of this result arises from that the Mittag-Leffler
function is well studied (see, e.g., Ref.~\cite{HMS} and references
therein). In particular, using the series definition of this
function, $E_{\alpha,\beta}(z) = \sum_{n=0}^{\infty} z^{n}/ \Gamma
(\alpha n + \beta)$, we obtain
\begin{equation}
    \mathcal{P}(y) = \frac{H(\sigma y)}{|y|^{1-\alpha}}
    \sum_{n=0}^{\infty} \frac{(-1)^{n} |y|^{\alpha n}}
    {\Gamma[\alpha(n+1)]}.
\label{limP_M-L1}
\end{equation}
Remarkably, the limiting probability density (\ref{limP_M-L}) at
$\alpha = 1/2$ can be expressed through a simple complementary error
function $\mathrm{erfc}(z) = (2/\sqrt{\pi}) \int_{z}^{\infty} dx
e^{-x^{2}}$. To show this, we first note that
\begin{equation}
    E_{\alpha,\alpha}(-|y|^{\alpha}) = -|y|^{1-\alpha}
    \frac{d}{d|y|}E_{\alpha,1}(-|y|^{\alpha}).
\label{relat10}
\end{equation}
(This relation follows directly from the series representation of the
Mittag-Leffler function.) Then, using the known result \cite{HMS}
$E_{1/2,1}(-z) = e^{z^{2}} \mathrm{erfc}(z)$, Eq.~(\ref{limP_M-L}) at
$\alpha = 1/2$ can be written in the form
\begin{equation}
    \mathcal{P}(y) = H(\sigma y)\left[ \frac{1}
    {\sqrt{\pi |y|}} - e^{|y|} \mathrm{erfc}\big(
    \sqrt{|y|}\big)\right]\! .
\label{P_1/2}
\end{equation}
The plot of this density function is shown in Fig.~\ref{fig1}.

\subsection{Series representation of $\bm{\mathcal{P} (y)}$}

We complete our study of alternative forms of the limiting
probability density $\mathcal{P}(y)$ by determining its series
representation. This representation can be useful, for example, for
the numerical evaluation of $\mathcal{P}(y)$, especially in the
vicinity of small $|y|$. Our starting point is the limiting
probability density written in the form of inverse Mellin transform
$\mathcal{P}(y) = (2 \pi i)^{-1} \int_{c- i\infty}^{c+ \infty} d\eta
S(\eta)$ with
\begin{equation}
    S(\eta) = \Gamma(1 - \alpha \eta)
    \frac{\sin[\phi(y) \eta]}{\sin(\pi \eta)}
    |y|^{\alpha \eta -1}
\label{def_S}
\end{equation}
[see the first line of Eq.~(\ref{limP_tr})]. To calculate the above
integral, we close the integration path by a semicircle $C_{R}$ of a
large radius $R$, which lies in the right half-plane of the complex
variable $\eta$. If this semicircle does not cross any singularity of
$S(\eta)$, then, using the Stirling approximation for the $\Gamma$
function \cite{Erd2}, it can be shown that the contribution of
$C_{R}$ into the integral over the closed contour $L$ vanishes as $R
\to \infty$. Therefore, from the residue theorem (see, e.g.,
Ref.~\cite{AF}) we obtain $\mathcal{P}(y) = - \sum_{j}
\mathrm{Res}(S, \eta_{j})$, where $\mathrm{Res}(S, \eta_{j})$ denotes
the residue of $S(\eta)$ at $\eta = \eta_{j}$, the sum is taken over
all isolated singularities (in our case poles) of $S(\eta)$ inside
the contour $L$, and the sign $``-"$ accounts for the direction of
$L$.

According to Eq.~(\ref{def_S}), the poles of $S(\eta)$ result from
the first-order poles $\eta_{n} = n/\alpha$ ($n \geq 1$) of $\Gamma(1
- \alpha \eta)$ and from the first-order poles $\eta_{m} = m$ ($m
\geq 1$) of $1/ \sin(\pi \eta)$. If $\alpha$ is irrational, then
these sets of poles, $\{\eta_{n}\}$ and $\{\eta_{m}\}$, are not
intersected, and hence all poles of $S(\eta)$ are also first-order.
However, if $\alpha$ is rational, then some (or all if $\alpha=1$)
poles from the set $\{\eta_{n}\}$ coincide with some (or all) poles
from the set $\{\eta_{m}\}$, resulting in the appearance of the
second-order poles of $S(\eta)$. Since the cases with irrational and
rational values of $\alpha$ seem quite different, we consider them
separately.

\subsubsection{Irrational values of $\alpha$}

In this case, the limiting probability density is written as
$\mathcal{P}(y) = - \sum_{n=1}^{\infty} \mathrm{Res}(S, n/\alpha) -
\sum_{m=1}^{\infty} \mathrm{Res}(S, m)$. Therefore, taking into
account that $\Gamma(1- \alpha\eta)|_{\eta = n/\alpha + \xi} \sim
(-1)^{n}/ [\alpha \Gamma(n)\xi]$ and $1/\sin(\pi \eta)|_{\eta=m+ \xi}
\sim (-1)^{m}/ (\pi \xi)$ as $\xi \to 0$, and using the reflection
formula $\Gamma(1- \alpha m) = \pi/ [\Gamma(\alpha m) \sin(\pi \alpha
m)]$, we readily find
\begin{eqnarray}
    \mathcal{P}(y) \!\!&=&\!\! \frac{1}{\alpha}
    \sum_{n=1}^{\infty} \frac{(-1)^{n-1}
    \sin[\phi(y)n/\alpha]} {\Gamma(n)
    \sin(\pi n/\alpha)}|y|^{n -1}
    \nonumber\\ [4pt]
    &&\!\! + \sum_{m=1}^{\infty}\frac{(-1)^{m-1}
    \sin[\phi(y)m]}{\Gamma(\alpha m)\sin(\pi \alpha m)}
    |y|^{\alpha m -1}. \quad
\label{limP_ser1}
\end{eqnarray}

\subsubsection{Rational values of $\alpha$}

Let us assume now that the tail parameter $\alpha$ is given by the
irreducible fraction $\alpha = l/p$, where $l(\geq 1)$ and $p (\geq
1)$ are natural numbers satisfying the condition $l \leq 2p$. In this
case, the first-order poles of $\Gamma(1 - p \eta /l)$ with numbers
$n=lk$ ($k=1,2,\ldots$) and the first-order poles of $1/ \sin(\pi
\eta)$ with numbers $m=pk$ are merged, and thus the poles of
$S(\eta)$ at $\eta = pk$ become second-order. It is therefore
convenient to represent the limiting probability density in the form
\begin{eqnarray}
    \mathcal{P}(y) \!\!&=&\!\! -
    \mspace{-20mu} \sum_{\substack{ n=1 \\
    (n \neq l, 2l,\ldots)}}^{\infty}
    \mspace{-20mu} \mathrm{Res}(S, pn/l) -
    \mspace{-20mu} \sum_{\substack{ m=1 \\
    (m \neq p, 2p,\ldots)}}^{\infty}
    \mspace{-20mu} \mathrm{Res}(S, m)
    \nonumber\\
    &&\!\! - \sum_{k=1}^{\infty}\mathrm{Res}(S, pk),
\label{limP_res}
\end{eqnarray}
where the last sum is over all second-order poles of $S(k)$. Using
the above results for the residues of $S(\eta)$ at the first-order
poles and the asymptotic formula \cite{Erd2} $\Gamma(1- l\eta/p)
|_{\eta = pk + \xi} \sim (-1)^{lk}(p/l)[\xi^{-1} - (l/p)
\psi(lk)]/\Gamma(lk)$ ($\xi \to 0$) with $\psi(x) = d\ln\Gamma
(x)/dx$ being the $\psi$ (or digamma) function, from
Eq.~(\ref{limP_res}) we obtain
\begin{eqnarray}
    \mathcal{P}(y) \!\!&=&\!\! \frac{p}{l}
    \mspace{-20mu} \sum_{\substack{ n=1 \\
    (n \neq l, 2l,\ldots)}}^{\infty}
    \mspace{-20mu} \frac{(-1)^{n-1}
    \sin[\phi(y)pn/l]}{\Gamma(n)
    \sin(\pi pn/l)}|y|^{n -1}
    \nonumber\\
    &&\!\! + \mspace{-20mu} \sum_{\substack{ m=1 \\
    (m \neq p, 2p,\ldots)}}^{\infty}
    \mspace{-24mu} \frac{(-1)^{m-1}
    \sin[\phi(y)m]}{\Gamma(l m/p)
    \sin(\pi lm/p)} |y|^{lm/p -1}
    \nonumber\\
    &&\!\! +\, \frac{1}{\pi} \sum_{k=1}^{\infty}
    \frac{(-1)^{pk+lk}}{\Gamma(lk)}
    \Big( [\psi(lk) - \ln |y|] \sin[\phi(y)pk]
    \nonumber\\[4pt]
    &&\!\! -\, \frac{p}{l}\phi(y)\cos[\phi(y)pk]\Big)
    |y|^{lk-1}.
\label{limP_ser2}
\end{eqnarray}

It should be noted that if $\alpha \in (0,1)$ and $\alpha_{+} \neq
\alpha_{-}$ then the function $\phi(y)$ is given by Eq.~(\ref{psi1}).
In this case Eqs.~(\ref{limP_ser1}) and (\ref{limP_ser2}) are reduced
to Eq.~(\ref{limP_M-L1}) leading to the Mittag-Leffler function
representation (\ref{limP_M-L}). If $\alpha \in (1,2)$ and
$\alpha_{+} \neq \alpha_{-}$ then $\mathcal{P}(y)$ can also be
expressed in terms of the Mittag-Leffler function. Indeed, using
Eq.~(\ref{psi2}) and the relations $\sum_{n=0}^{\infty} (\pm
1)^{n}|y|^{n}/n! = e^{\pm|y|}$ and $\sum_{n=1}^{\infty} |y|^{\alpha
n}/\Gamma(\alpha n) = E_{\alpha, 0} (|y|^{\alpha}) = |y|^{\alpha}
E_{\alpha, \alpha} (|y|^{\alpha})$, Eqs.~(\ref{limP_ser1}) and
(\ref{limP_ser2}) can easily be reduced to
\begin{eqnarray}
    \mathcal{P}(y) \!\!&=&\!\! H(\sigma y)\bigg(
    \frac{e^{|y|}}{\alpha} -|y|^{\alpha-1}
    E_{\alpha, \alpha}(|y|^{\alpha}) \bigg)
    \nonumber\\ [4pt]
    &&\!\! +\, H(-\sigma y) \frac{e^{-|y|}}
    {\alpha}.
\label{limP_M-L2}
\end{eqnarray}

\section{SHORT- AND LONG-DISTANCE BEHAVIOR OF $\bm{\mathcal{P}
(y)}$} \label{beh}

The short-distance behavior of the limiting probability density
$\mathcal{P}(y)$ is completely described by the series
representations (\ref{limP_ser1}) and (\ref{limP_ser2}). To find the
long-distance behavior of $\mathcal{P}(y)$, it is convenient to use
its Laplace transform representation (\ref{limP_L}). According to
Watson's lemma \cite{AF}, the asymptotic series expansion of
$\mathcal{P}(y)$ at $|y| \to \infty$ is determined from the small-$x$
series expansion of the integrand function multiplied by $e^{|y|x}$.
Therefore, using the series expansion \cite{Dw}
\begin{equation}
    \frac{\sin[\phi(y)] x^{\alpha}}{1 +
    2\cos[\phi(y)]x^{\alpha} + x^{2\alpha}} =
    \sum_{n=1}^{\infty} (-1)^{n-1}\sin[\phi(y)n]
    x^{\alpha n}
\label{ser1}
\end{equation}
($|x|<1$) and the standard integral \cite{PBM}
\begin{equation}
    \int_{0}^{\infty} dx e^{-|y|x} x^{\alpha n} =
    \frac{\Gamma(1+\alpha n)}{|y|^{1+\alpha n}},
\label{int4}
\end{equation}
from Eq.~(\ref{limP_L}) one obtains
\begin{equation}
    \mathcal{P}(y) \sim \frac{1}{\pi}
    \sum_{n=1}^{\infty} (-1)^{n-1}\sin[\phi(y)n]
    \frac{\Gamma(1+\alpha n)}{|y|^{1+\alpha n}}
\label{as_P}
\end{equation}
as $|y| \to \infty$. For all cases of interest, we list below the
main terms of $\mathcal{P}(y)$ at $|y| \to 0$ and $|y| \to \infty$.

\subsubsection*{1. $\;\,$ $\alpha \in (0,1)$, $\alpha_{+}
\neq \alpha_{-}$}

In this case, the two-valued function $\phi(y)$ is given by
Eq.~(\ref{psi1}), $\mathcal{P}(y)|_{\sigma y<0}=0$, and thus
Eqs.~(\ref{limP_M-L1}) and (\ref{as_P}) lead to
\begin{equation}
    \mathcal{P}(y)|_{\sigma y>0} \sim \frac{1}
    {\Gamma(\alpha)} \frac{1}{|y|^{1-\alpha}}
\label{P0_1}
\end{equation}
($|y| \to 0$) and to
\begin{equation}
    \mathcal{P}(y)|_{\sigma y>0} \sim \frac{1}
    {\pi} \sin(\pi\alpha)\Gamma(1+\alpha)
    \frac{1}{|y|^{1+\alpha}}
\label{Pas_1}
\end{equation}
($|y| \to \infty$), respectively.

\subsubsection*{2. $\;\,$ $\alpha \in (1,2)$, $\alpha_{+}
\neq \alpha_{-}$, $l_{1}=0$}

For these conditions, the function $\phi(y)$ is determined from
Eq.~(\ref{psi2}), $\mathcal{P}(y)|_{\sigma y<0}=e^{-|y|}/\alpha$, and
so Eqs.~(\ref{limP_ser1}), (\ref{limP_ser2}) and (\ref{as_P}) yield
\begin{equation}
    \mathcal{P}(y)|_{\sigma y>0} \sim
    \frac{1}{\alpha} - \frac{1}{\Gamma(\alpha)}
    |y|^{\alpha-1}
\label{P0_2}
\end{equation}
as $|y| \to 0$ and
\begin{equation}
    \mathcal{P}(y)|_{\sigma y>0} \sim -
    \frac{1}{\pi}\sin(\pi\alpha)\Gamma(1+\alpha)
    \frac{1}{|y|^{1+\alpha}}
\label{Pas_2}
\end{equation}
as $|y| \to \infty$.

\subsubsection*{3. $\;\,$ $\alpha_{+} = \alpha_{-} =
\alpha \in (0,1)$, $u_{+} \neq u_{-}$}

Using Eq.~(\ref{psi3}), it can be straightforwardly shown that
\begin{equation}
    \mathcal{P}(y) \sim \frac{1 + \mathrm{sgn}
    (\epsilon y)|\epsilon|}{2\Gamma(\alpha)
    \sqrt{\epsilon^{2} + (1-\epsilon^{2})
    \cos^{2} (\pi \alpha/2)}}\frac{1}{|y|^{1-\alpha}}
\label{P0_3}
\end{equation}
as $|y| \to 0$ and
\begin{equation}
    \mathcal{P}(y) \sim \frac{[1 + \mathrm{sgn}
    (\epsilon y)|\epsilon|]\sin(\pi\alpha)
    \Gamma(1+\alpha)}{2\pi\sqrt{\epsilon^{2} +
    (1-\epsilon^{2})\cos^{2} (\pi \alpha/2)}}
    \frac{1}{|y|^{1+\alpha}}
\label{Pas_3}
\end{equation}
as $|y| \to \infty$. In contrast to the first case, the limiting
probability density has both left and right branches characterized by
the same tail index $\alpha$.

\subsubsection*{4. $\;\,$ $\alpha_{+} = \alpha_{-} =
\alpha \in (1,2)$, $u_{+} \neq u_{-}$, $l_{1}=0$}

Finally, in this case we have
\begin{equation}
    \mathcal{P}(y) \sim \mathcal{P}(0) - \frac{1 +
    \mathrm{sgn}(\epsilon y)|\epsilon|}{2\Gamma(\alpha)
    \sqrt{\epsilon^{2} + (1-\epsilon^{2})
    \cos^{2} (\pi \alpha/2)}}|y|^{\alpha -1}
\label{P0_4}
\end{equation}
as $|y| \to 0$ and
\begin{equation}
    \mathcal{P}(y) \sim -\frac{[1 + \mathrm{sgn}
    (\epsilon y)|\epsilon|]\sin(\pi\alpha)
    \Gamma(1+\alpha)}{2\pi \sqrt{\epsilon^{2} +
    (1-\epsilon^{2})\cos^{2} (\pi \alpha/2)}}
    \frac{1}{|y|^{1+\alpha}}
\label{Pas_4}
\end{equation}
as $|y| \to \infty$, where $\mathcal{P}(0)$ is given by
Eq.~(\ref{P(0)2}). Note that at $\epsilon =0$
Eqs.~(\ref{P0_3})--(\ref{Pas_4}) are reduced to those obtained in
Ref.~\cite{DYBKL} for symmetric walks.

\section{NUMERICAL SIMULATION OF $\bm{\mathcal{P} (y)}$}
\label{Num}

The determination of the limiting probability density
$\mathcal{P}(y)$ by the numerical simulation is not a trivial
problem. To understand why this is so, we first recall that
$\mathcal{P} (y)$ is the probability density of the random variable
$Y(t) = a(t)X(t)$ in the limit $t \to \infty$. In the simulation,
however, we need to consider the behavior of the variable $Y(T) =
a(T)X(T)$ and the corresponding probability density
\begin{equation}
    \mathcal{P}_{T}(y) = \frac{1}{a(T)} \,P \!
    \left( \frac{y}{a(T)}, T \right)
\label{P_T}
\end{equation}
for some finite time $t=T$. To be sure that this probability density
approaches $\mathcal{P}(y)$, the operating time $T$ must be large
enough and, in principle, it should exceed the characteristic scale
of waiting times. But in our case all fractional moments of the
waiting-time density $p(\tau)$ do not exist and thus there is no
finite time scale of $p(\tau)$. This means that for any finite $T$
there is always a non-negligible survival probability $V(T) =
\int_{T}^ {\infty} d\tau p(\tau)$ that the waiting time is larger
than $T$. Therefore, the minimal value of $T$ is restricted only by
the condition $a(T)\ll 1$ which is equivalent to $V(T)\ll 1$. Since
$V(T)$ is a slowly varying function, it decreases with increasing of
$T$ very slowly, and thus the operating time is expected to be very
large. On the other hand, the larger is $T$ the larger is the average
number $\overline{N}(T)$ of jumps occurring in the time interval
$(0,T)$ [this is so because, according to \cite{Hug}, $\overline
{N}(T) \sim V^{-1}(T)$], and hence the larger is the computational
time. Thus, the chosen value of the operating time $T$ must satisfy
the condition $V(T) \ll 1$ and provide a reasonable computational
time.

In our numerical simulations, we use the following waiting-time
probability density:
\begin{equation}
    p(\tau) = \frac{v \ln^{v}g}{(g + \tau)
    \ln^{1+v}(g + \tau)}
\label{p}
\end{equation}
with $v>0$ and $g>1$. The main advantage of this density is that its
distribution function $F_{p}(\tau) = \int_{0}^{\tau} d\tau'p(\tau')$
is calculated explicitly
\begin{equation}
    F_{p}(\tau) = 1- \frac{\ln^{v}g}{\ln^{v}(g + \tau)},
\label{F_p}
\end{equation}
and thus the inverse function of $U = F_{p}(\tau)$ is given by $\tau
= g^{(1-U)^{-1/v}} - g$. The last result permits us to use the
inversion method \cite{Dev} in accordance with which the random
variables defined as
\begin{equation}
    \tau_{n} = g^{(1-U_{n})^{-1/v}} - g,
\label{tau_n}
\end{equation}
where $n=1,2,\ldots$ and $U_{n}$ are random numbers uniformly
distributed in the interval $[0,1]$, have the same probability
density (\ref{p}). This provides a simple way to generate the waiting
times. For the simulations we choose $g=2, v=2$ and $T=10^{15}$
yielding $V(T)  \approx 4\cdot 10^{-4} $.

Our choice of the jump density $w(\xi)$ is limited by two conditions.
First, to verify the theoretical results, $w(\xi)$ must reproduce all
possible cases considered earlier and, second, to simplify the
generation of the jump lengths $\xi_{n}$, the corresponding
distribution function $F_{w}(\xi) = \int_{-\infty}^{ \xi} d\xi'
w(\xi')$ must be invertible. These conditions are satisfied, for
example, by the jump density
\begin{equation}
    w(\xi) = \left\{\!\! \begin{array}{cl}
    \alpha_{-} c_{-}b_{-}^{\alpha_{-}}/
    (b_{-} - \xi)^{1+\alpha_{-}},
    & \xi < 0
    \\ [6pt]
    \alpha_{+} c_{+}b_{+}^{\alpha_{+}}/
    (b_{+} + \xi)^{1+\alpha_{+}},
    & \xi \geq 0.
    \end{array}
    \right.
\label{w}
\end{equation}
Here, $b_{\pm} \in (0, \infty)$ and $c_{+} + c_{-} =1$ with $c_{+}$
and $c_{-}$ being the probabilities that $\xi \geq 0$ and $\xi<0$,
respectively. It can be easily shown from Eq.~(\ref{w}) that
\begin{equation}
    F_{w}(\xi) = \left\{\!\! \begin{array}{cl}
    c_{-}b_{-}^{\alpha_{-}}/
    (b_{-} - \xi)^{\alpha_{-}},
    & \xi < 0
    \\ [6pt]
    1 - c_{+}b_{+}^{\alpha_{+}}/
    (b_{+} + \xi)^{\alpha_{+}},
    & \xi \geq 0,
    \end{array}
    \right.
\label{F_w}
\end{equation}
and thus the jump lengths can be determined as
\begin{equation}
    \xi_{n} = \left\{\!\! \begin{array}{cl}
    -b_{-}(c_{-}/U_{n})^{1/\alpha_{-}} + b_{-},
    & U_{n} < c_{-}
    \\ [6pt]
    b_{+}[c_{+}/(1-U_{n})]^{1/\alpha_{+}}-b_{+},
    & U_{n} \geq c_{-}.
    \end{array}
    \right.
\label{xi_n}
\end{equation}

Now we are ready to describe the procedure for calculating
$\mathcal{P}_{T}(y)$. According to the definition, the particle
starts to walk at time $t = 0$ from the position $X(0) = 0$. At the
first step, the waiting time $\tau_{1}$ and the jump length $\xi_{1}$
are generated using Eqs.~(\ref{tau_n}) and (\ref{xi_n}),
respectively. If $\tau_{1}\leq T$, then the particle position becomes
$X(\tau_{1}) = \xi_{1}$, and we can go to second step that consists
in generating new random numbers $\tau_{2}$ and $\xi_{2}$. If at the
$n$th step the condition $\sum_{i=1}^{n} \tau_{i} \leq T$ is
violated, then the walk is stopped at the previous step, i.e., the
scaled position $Y(t)$ of the first particle at $t=T$ is assumed to
be $Y(T) = a(T)\sum_{i=1}^{n-1} \xi_{i}$ [$Y(T) =0$ at $n=1$], where
the scaling function $a(T)$ can be calculated using an appropriate
theoretical formula. Determining $Y(T)$ for $N$ particles, we can
evaluate the limiting probability density as follows:
$\mathcal{P}_{T}(y) = N_{\Delta y}/N$, where $N_{\Delta y}$ is the
number of particles with $Y(T) \in [y,y+\Delta y)$. In all our
simulations we set $N=10^{5}$ and $\Delta y = 10^{-1}$; the other
parameters are listed below.

\subsubsection*{1. $\;\,$ $\alpha \in (0,1)$, $\alpha_{+}
\neq \alpha_{-}$}

In this case, the limiting probability density $\mathcal{P}(y)$
depends only on the minimal tail index $\alpha$ [see, e.g.,
Eq.~(\ref{limP_M1})]. But to determine the approximate probability
density $\mathcal{P}_{T}(y)$ by the proposed procedure, all the
parameters in Eqs.~(\ref{p}) and (\ref{w}) must be specified. In
addition to the parameters mentioned above, we choose $\alpha
=\alpha_{+} = 1/2$, $c_{+} = 2/3$, $b_{+} =1$ and $\alpha_{-} = 3/4$,
$c_{-} = 1/3$, $b_{-} =1$. With these parameters, Eq.~(\ref{as_a3})
yields $a(T) \approx 1.2\cdot 10^{-7}$ and the simulated values of
$\mathcal{P}_{T}(y)$, marked by red triangles, are shown in
Fig.~\ref{fig1}.

\subsubsection*{2. $\;\,$ $\alpha \in (1,2)$, $\alpha_{+}
\neq \alpha_{-}$, $l_{1}=0$}

Since the first moment of the probability density (\ref{w}) is given
by
\begin{equation}
    l_{1} = \frac{c_{+}b_{+}}{\alpha_{+} -1} -
    \frac{c_{-}b_{-}}{\alpha_{-} -1},
\label{l1}
\end{equation}
($\alpha_{\pm}>1$) the condition $l_{1}=0$ implies that the
parameters of $w(\xi)$ satisfy the condition $c_{+}b_{+}/ (\alpha_{+}
-1) = c_{-}b_{-}/ (\alpha_{-} -1)$. In particular, if $\alpha
=\alpha_{+} = 5/4$, $c_{+} = 5/22$, $b_{+} =1$ and $\alpha_{-} =
37/20$, $c_{-} = 17/22$, $b_{-} =1$, then $c_{+}b_{+}/ (\alpha_{+}
-1) = c_{-}b_{-}/ (\alpha_{-} -1) = 10/11$, $a(T) \approx 1.8\cdot
10^{-3}$, and the results of simulation of $\mathcal{P}_{T}(y)$ are
shown in Fig.~\ref{fig2}.

\subsubsection*{3. $\;\,$ $\alpha_{+} = \alpha_{-} =
\alpha \in (0,1)$, $u_{+} \neq u_{-}$}

According to Eq.~(\ref{w}) and the asymptotic formula
(\ref{w_asymm}), the parameters $u_{\pm}$ are expressed through the
parameters of $w(\xi)$ as follows: $u_{\pm} = \alpha_{\pm} c_{\pm}
b_{\pm}^{\alpha_{\pm}}$. Keeping in mind that $\alpha_{+} =
\alpha_{-} = \alpha$ and $u_{+} \neq u_{-}$, i.e., $c_{+} b_{+}^{
\alpha_{+}} \neq c_{-} b_{-}^{\alpha_{-}}$, we choose $\alpha = 1/2$,
$c_{+} = 2/3$, $b_{+} =1$ and $c_{-} = 1/3$, $b_{-} =1$. For these
parameters, $u_{+} = 1/3$, $u_{-} = 1/6$ (i.e., $\epsilon = 1/3$),
$a(T) \approx 9.3 \cdot 10^{-8}$ and the simulated results are shown
in Fig.~\ref{fig3}.

\subsubsection*{4. $\;\,$ $\alpha_{+} = \alpha_{-} =
\alpha \in (1,2)$, $u_{+} \neq u_{-}$, $l_{1}=0$}

In this case, the conditions $u_{+} \neq u_{-}$ and $l_{1}=0$ are
reduced to $c_{+} b_{+}^{ \alpha} \neq c_{-} b_{-}^{\alpha}$ and
$c_{+}b_{+} = c_{-}b_{-}$, respectively. Choosing $\alpha = 5/4$,
$c_{+} = 1/5$, $b_{+} =5$ and $c_{-} = 4/5$, $b_{-} =5/4$, one gets
$u_{+} = 5^{5/4}/4 \approx 1.87$, $u_{-}= (5/4)^{5/4} \approx 1.32$
[i.e., $\epsilon = (\sqrt{2} -1)/(\sqrt{2} +1) \approx 0.17$], $a(T)
\approx 5.2 \cdot 10^{-4}$ and the simulated values of
$\mathcal{P}_{T}(y)$ are shown in Fig.~\ref{fig4}.

As is seen from these figures, the numerical results are in very good
agreement with our theoretical predictions. It is also worth to note
that the proposed numerical method reproduces all the other limiting
probability densities, Eqs.~(\ref{limP3}), (\ref{limP5}) and
(\ref{limP7}), and can easily be extended to study the CTRWs,
including coupled ones, in higher dimensions.

\section{CONCLUSIONS}
\label{Concl}

We have studied in detail the long-time behavior of the decoupled
CTRWs characterized by superheavy-tailed distributions of waiting
times and asymmetric heavy-tailed distributions of jump lengths. The
main attention is devoted to introducing the scaled particle position
and deriving its limiting probability density $\mathcal{P}(y)$. Using
the Montroll-Weiss equation in the Fourier-Laplace space and the
asymptotic properties of the waiting-time and jump-length
distributions, we have found both the scaling function, which
determines the scaled position, and the representation of
$\mathcal{P}(y)$ in terms of the inverse Fourier transform. It has
been shown that while the scaling function depends on the parameters
describing the asymptotic behavior of both waiting-time and
jump-length distributions, the limiting probability density is
completely characterized by the parameters of the latter
distribution. Among these parameters, the main role plays the
smallest tail index $\alpha$.

To get more information about the limiting probability density, we
have derived a number of alternative representations of
$\mathcal{P}(y)$. The representation of $\mathcal{P}(y)$ in terms of
the inverse Mellin transform is important from a theoretical point of
view (all other representations considered in the paper follow from
this one) and permits to determine the intervals of $\alpha$ where
$\mathcal{P}(y)$ exhibits qualitatively different behavior. We have
also obtained the limiting probability density in terms of the Fox
$H$ function. The importance of this representation is that the $H$
function is well studied and many of the special functions can be
considered as its particular cases. In particular, we have shown that
if the tail indexes of the jump density are different and $\alpha \in
(0,1)$ then $\mathcal{P}(y)$ is expressed through the generalized
Mittag-Leffler function. Then, using the limiting density in terms of
the Laplace transform, we have analytically demonstrated that
$\mathcal{P}(y)$ is non-negative, has a maximum value at the origin,
and monotonically decreases (or equals zero) as $|y|$ increases. We
have also derived the series representation of $\mathcal{P}(y)$
which, in the case when the tail indexes of the jump density are
different and $\alpha \in (1,2)$, has been used to obtain the
limiting density in terms of the Mittag-Leffler function. Moreover,
the series representation of $\mathcal{P}(y)$ together with its
Laplace transform representation have permitted us to completely
describe the short- and long-distance behavior of the limiting
probability density. It has been shown, in particular, that the tail
index of $\mathcal{P}(y)$ is equal to the lowest tail index of the
jump probability density.

Finally, we have developed a numerical method for calculating the
limiting probability density. This method, which deals with the
statistics of the scaled particle position at large times, has been
applied to calculate $\mathcal{P}(y)$ in all cases of interest. It
has been shown that the simulation results for the limiting
probability density are in excellent agreement with our theoretical
predictions.

\section*{ACKNOWLEDGMENTS}

S.I.D.~and Yu.S.B.~are grateful to the Ministry of Education and
Science of Ukraine for financial support under grant No.~0112U001383
and under the mobility program (Order No.~650 of 31.05.2012)
(S.I.D.).

\appendix*

\section{Fractional equation for $\bm{\mathcal{P} (y)}$}
\label{frac}

Let us define the Riesz-Feller space-fractional derivative ${}_{y}D_{
\theta}^{\gamma}$ of order $\gamma$ and skewness $\theta$ as (see,
e.g., Ref.~\cite{MLP})
\begin{equation}
    \mathcal{F}\{{}_{y}D_{\theta}^{\gamma} f(y)\} =
    -e^{i\mathrm{sgn}(\kappa)\pi \theta/2}
    |\kappa|^{\gamma}f_{\kappa},
\label{defFrac}
\end{equation}
where $\gamma \in (0,2]$, $|\theta| \leq \min\{\gamma, 2-\gamma\}$,
and $\mathcal{F}\{f(y)\} = \int_{-\infty}^{\infty} dy e^{i\kappa y}
f(y) = f_{\kappa}$. Then, using this definition and Eq.~(\ref{Phi3}),
one gets
\begin{eqnarray}
    \mathcal{F}\{{}_{y}D_{-2\varphi/\pi}^{\alpha}
    \mathcal{P}(y)\}  \!\!&=&\!\! -e^{-i\mathrm{sgn}
    (\kappa)\varphi} |\kappa|^{\alpha}\mathcal{P}_{\kappa}
    \nonumber\\ [4pt]
    \!\!&=&\!\! - \Phi(\kappa) \mathcal{P}_{\kappa}.
\label{rel12}
\end{eqnarray}
Finally, taking into account that $\mathcal{F} \{\delta(y)\} =1$ and,
as it follows from Eq.~(\ref{limP2}), $\mathcal{F} \{\mathcal{P}
(y)\} =[1+\Phi(\kappa)]^{-1}$, the fractional equation for the
limiting probability density $\mathcal{P}(y)$ can be written in the
form
\begin{equation}
    {}_{y}D_{-2\varphi/\pi}^{\alpha} \mathcal{P}(y) =
    \mathcal{P}(y) - \delta(y).
\label{FracEq}
\end{equation}

\end{document}